\documentclass[pdflatex,sn-mathphys-num]{sn-jnl}

\usepackage{graphicx}%
\usepackage{multirow}%
\usepackage{amsmath,amssymb,amsfonts}%
\usepackage{amsthm}%
\usepackage{mathrsfs}%
\usepackage[title]{appendix}%
\usepackage{xcolor}%
\usepackage{textcomp}%
\usepackage{manyfoot}%
\usepackage{booktabs}%
\usepackage{algorithm}%
\usepackage{algorithmicx}%
\usepackage{algpseudocode}%
\usepackage{listings}%

\usepackage{color}
\usepackage{multirow}
\usepackage{subfigure}
\usepackage[export]{adjustbox}
\usepackage{enumitem}
\usepackage{subcaption}
\usepackage{hhline}
\usepackage{balance}
\usepackage{colortbl}
\usepackage{units}
\usepackage{marginnote}
\usepackage{pgfplotstable,array}
\usepackage{marginnote}
\usepackage{float}
\usepackage{pgfplots}
\usepackage{pgf,tikz}
\usepackage{mathtools}
\usepackage{xstring}
\usepackage{pifont}
\usepackage{wrapfig}
\usepackage{amsmath}

\def\checkmark{\tikz\fill[scale=0.4](0,.35) -- (.25,0) -- (1,.7) -- (.25,.15) -- cycle;} 

\newcommand{\xmark}{\ding{55}}

\newcommand{\stitle}[1]{\vspace{0.8ex}\noindent\textup{\textbf{#1}}}

\usepackage{algorithmicx}
\usepackage{algorithm}
\usepackage{algpseudocode}

\newtheorem{lemma}{Lemma}

\newcommand{\hide}[1]{}

\AtBeginDocument{%
  }

\theoremstyle{thmstyleone}%
\theoremstyle{thmstyletwo}%

\theoremstyle{thmstylethree}%
\newtheorem{definition}{Definition}%

\begin{document}

\title[Article Title]{Multi-granularity Spatiotemporal Flow Patterns}


\author*[1]{\sur{Chrysanthi Kosyfaki}}\email{ckosyfaki@cse.ust.hk}

\author[2]{\sur{Nikos Mamoulis}}\email{nikos@cs.uoi.gr}

\author[3]{\sur{Reynold Cheng}}\email{ckcheng@cs.hku.hk}

\author[3]{\sur{Ben Kao}}\email{kao@cs.hku.hk}

\affil*[1]{\orgdiv{Department of Computer Science and Engineering},
  \orgname{Hong Kong University of Science and Technology}, \city{Hong Kong},
    \country{China}}

\affil[2]{\orgdiv{Department of Computer Science and Engineering},
  \orgname{University of Ioannina}, \city{Ioannina}, \country{Greece}}

\affil[3]{\orgdiv{Department of Computer Science}, \orgname{University
  of Hong Kong}, \city{Hong Kong}, \country{China}}


\abstract{Analyzing flow of objects or data at
    different granularities of space and time can unveil interesting
    insights or trends. For example, transportation companies, by
    aggregating passenger travel data (e.g., counting passengers
    traveling from one region to another), can analyze movement
    behavior. In this paper, we study the problem of finding important
  trends in passenger movements between regions at different
  granularities. We define Origin ($O$), Destination ($D$), and Time
  ($T$) patterns (ODT patterns) and propose a bottom-up algorithm that
  enumerates them. We suggest and employ optimizations that greatly reduce the
  search space and the computational cost of pattern enumeration. We
  also propose pattern variants (constrained patterns and top-$k$
  patterns) that could be useful to different applications
  scenarios. Finally, we propose an approximate solution that
  fast identifies ODT patterns of specific sizes, following a
  generate-and-test approach.
  We evaluate the efficiency and effectiveness of
  our methods on three real datasets and
  showcase interesting ODT flow patterns in them.}






\maketitle

\section{Introduction}\label{sec:intro}
In a transportation network,
nodes represent stations (or districts)
and edges
model the connections between the nodes carrying 
information related to movement or flow (e.g., passengers travel 
between districts during different times of a day). Examples include
car or taxi passengers moving from one district of a city to another,
metro or railway passengers, global travellers by air, etc.
Transportation companies and
organizations
collect large volumes of data from their passengers,
regarding their origins, destinations, and times
of their trips.
For each individual trip, the available information is typically the
starting (or entry) location (or station) and time of the trip and the final
(or exit) location (or station) and time.
In-between locations (i.e., the entire trajectories) are often
unavailable due to incapability of tracing and privacy constraints.
For example, passengers of a metro network
use their travel or credit cards at the
entry and exit points of their trips and in-between locations can only
be inferred. Similarly, taxi companies
usually report the starting and ending locations (or districts)
of trips instead of the entire trajectories.

Information on individual trips can be used in
personalized services, after obtaining consent from the passengers.
Other than that, it is hard to use such detailed data, mainly due to
privacy constraints.
However, aggregate information about passenger trips can
be invaluable to the company since it can provide time-parameterized
estimates
and predictions about the passenger flow between regions,
which can help to
improve their service.

\begin{figure}[hbt]
    \centering
   \includegraphics[width=0.70\textwidth]{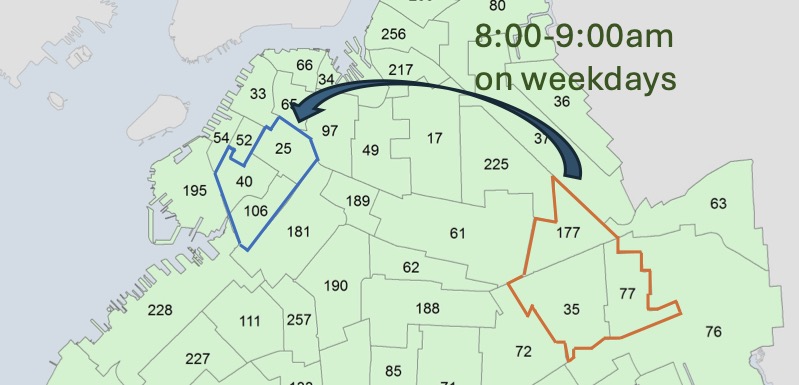}
     \caption{Example of an ODT pattern}
    \label{fig:brooklyn_pattern}
  \end{figure}

\stitle{Problem} We propose a novel analysis tool for
transportation data, called
\textit{Origin-Destination-Time} (ODT) patterns.
An ODT pattern captures
information about the origins, destinations, and
the volume (flow) of the passengers
at different times of the day where the trips take place.
We first use the application
domain to define the finest granularity of regions on the map (e.g.,
each region corresponds to a metro station or a district) and also
define the finest time intervals of interest (e.g., divide the
24-hour time interval of a day into 48 equal 30-minute timeslots or
use short slots at peak hours and long ones at night). We call
these \textit{atomic regions} and \textit{atomic timeslots},
respectively.
This means that, a trip can be described by its origin region, its destination
region, and the timeslot when the trip starts,
i.e., as an $(o,d,t)$ triple. 
For all $(o,d,t)$ triples, where $o$ and $d$ are (different) atomic
regions and $t$ is an atomic timeslot, we measure the total number of
passengers who took a trip from $o$ to $d$ at time $t$.
The total flow of an $(o,d,t)$ triple
characterizes its importance; the triples with high flow are
considered to be important and they are called {\em atomic
  ODT-patterns} (we drop the ODT prefix whenever the context is clear).
In a {\em generalized} ODT-triple, denoted by $(O,D,T)$, $O$ and $D$
are sets of {\em neighboring}
atomic regions and $T$ consists of one or more {\em consecutive} timeslots.
An atomic $(o,d,t)$ triple is a component of an  $(O,D,T)$ triple if
$o\in O$, $d\in D$, and $t\in T$. 
$(O,D,T)$ is {\em non-atomic}, if
at least one of $O$, $D$, or $T$ is non-atomic.
Figure \ref{fig:brooklyn_pattern} exemplifies a generalized
ODT pattern, which captures statistically significant movement in
Brooklyn from boroughs \{35,77,177\} to boroughs \{25,40,106\} in the
mornings of weekdays.

\stitle{Methodology}
The number of possible atomic region
combinations that can form a generalized (i.e., non-atomic) region $O$
or $D$ is huge and it is not practical to consider all these
combinations and their flows.
At the same time, for a given generalized
ODT triple, it is hard to estimate the flow quantity that can be deemed
significant enough to characterize the triple an interesting
pattern.
To solve these issues, we follow a {\em voting} approach, where we
characterize an ODT triple as a pattern if at least a certain percentage
of its constituent $(o,d,t)$ triples are atomic patterns (i.e., they
have large enough flow). This allows us to design and use a
pattern enumeration algorithm,
which, starting from the atomic patterns,
identifies all ODT patterns progressively by synthesizing them
from less generalized ODT patterns.
We propose a number of optimizations,
which
significantly reduce the time spent for generating candidate patterns
and measuring their significance.

\stitle{Problem variants}
Given the potentially large number of ODT patterns
and the high cost of their enumeration,
we also study practical ODT variants.
First, we investigate the detection of patterns which are {\em constrained} to a
subset of regions and timeslots, 
allowing us to focus on under-represented regions and to greatly reduce
the pattern enumeration cost.
Second, we study pattern detection by limiting the number of
atomic regions and timeslots that a pattern may have.
Finally, we define and solve the problem of finding the
top-ranked patterns at each granularity level, by
a novel algorithm that significantly outperforms the baseline approach
of finding all patterns at each level and then selecting the top ones.

\stitle{Approximate algorithms}
In this paper, our primary focus shifts in addressing the computational challenges of enumerating
ODT patterns \cite{DBLP:conf/ssd/KosyfakiMCK25}. As we mention earlier, enumerating all the possible ODT
triples makes exhaustive enumeration computational prohibitive. To
futher accelerate the enumeration of generalized ODT patterns, we
propose an approximate solution that significantly reduces both space
and time complexity compared to the baseline algorithm.
Assuming that we are interested in ODT patterns, where the O, D, and T
components have specific sizes (e.g., O consists of 4 neighboring
districts), we first generate thee pools of O-candidates, D-candidates
and T-candidates. O-candidates and D-candidates are connected
subgraphs in the region neighborhood graph, which have the requested
sizes. T-candidates are sequences of consecutive timeslots of the
requested size (e.g., 2-hour periods). 
Our approximate algorithm follows a generate-and-test approach, which
samples triples of ODT combinations from the candidates and counts
whether they include a sufficient number of atomic patterns.
To improve the effectiveness of our approximate algorithm, we also
propose a weighted version, which tests the most promising ODT
combinations to become ODT patterns.

\stitle{Applications} The problem of identifying ODT patterns
finds application in a wide variety of
real-world transportation networks \cite{DBLP:journals/tkde/LiuLJXDTZ23,DBLP:conf/kdd/WangYCW0019,
yu2021discovering}, such as road networks, railways,
etc. For example,
passenger movement patterns can
facilitate the handling of emergencies or incidents
by metro systems.
Consider 
an accident that happened in Hong Kong subway system in
December
2021\footnote{https://www.thestandard.com.hk/breaking-news/section/4/183861/(Video)-MTR-door-flew-off,-disrupting-peak-hour-service}.
Scheduled trips from some districts were cancelled and passengers had
to be served by other means (i.e., buses). ODT patterns could
help to schedule on-demand transportation for affected
passengers as they could indicate origin-destination region pairs
served by the disrupted route with heavy demand at the time of the day
when the incident happened.
In general, ODT patterns
can help transportation companies to improve their schedule
and in managing their available resources (e.g., fleet,
employees).
Another application is identifying the correlation
between map districts for performing target-marketing, cross-district
advertisements, or location planning.
For example, if there is a heavy passenger flow between districts that are
far from each other, the municipality may consider allocating new space
for housing or opening new services near the residencies of
passengers, to reduce traffic and improve the environmental conditions. 

\stitle{Novelty}
The input to our problem are not the entire trajectories of
passengers, which is typically not available to transportation
companies or organizations, as already discussed.
Instead, we assume that we only have access to the (anonymized) trips
of passengers, either individually or aggregated (e.g., the total
number of passengers who took a trip from a specific origin to a
specific destination at a specific time).
For example, metro companies only have access to the entry and exit
points of passengers, whereas public taxi trip data only include the
origin/destination districts and time of the trip,
but not the route. 
Hence, sequence or trajectory data mining methods are not applicable
to our problem. 
In addition, our goal is also very different compared to the
objective of sequence pattern mining. We study the
identification of frequent generalized ODT triples, where the origins,
destinations, and timeslots could be generalized to be a subgraph of
neighboring regions or a time interval. On the other hand, the goal
of sequence/trajectory mining methods is to find frequent
subsequences.
Table \ref{table:prevwork} summarizes representative works on trafflic flow
mining (TFM),
sequential and trajectory pattern mining (STM), 
origin-destination flow prediction (ODFP), and travel time
estimation (TTE) problems and their differences compared to our
work.
In summary, all works consider time, but ours is the only work that
(i) finds generalized ODT patterns, (ii) does not require in-between
points of trajectories, and (iii) considers flow in pattern detection.  
For a detailed comparison to previous work, see Sec. \ref{sec:relwork}.

\begin{table}[h!]
 \caption{Previous works}
\vspace{-0.38cm}
\centering
\scriptsize
\begin{tabular}{|c|c c c  c|}
 \hline
Previous works& \multicolumn{4}{c|}{Characteristics}\\ 
 \hline
&ODT Generalization &Time & Req. Traject. & Flow \\ 
 
TFM (\cite{DBLP:journals/tits/DuPWBWGLL20,DBLP:conf/kdd/FangLSX21,DBLP:journals/tkde/LiuLJXDTZ23,xie2020urban}) & \xmark &\checkmark &\xmark&\checkmark\\
 
STM (\cite{DBLP:conf/icde/AgrawalS95,DBLP:conf/gis/AlvaresBKMMV07,DBLP:conf/sigmod/NgC04,DBLP:journals/pvldb/FanZWT16,DBLP:conf/kdd/GiannottiNPP07,DBLP:reference/gis/GudmundssonLW08,DBLP:conf/ssd/KalnisMB05}) & \xmark & \checkmark&\checkmark &\xmark \\
 
ODFP (\cite{DBLP:journals/eswa/BeernaertsDLBW19,DBLP:journals/jits/BeharaBC22,duan2019prediction,kumarage2023hybrid,ros2022practical,DBLP:conf/kdd/WangYCW0019})& \xmark&\checkmark & \xmark&\checkmark \\

TTE (\cite{DBLP:journals/pvldb/JinLDW11,lin2023origin,DBLP:journals/pvldb/00020B22,DBLP:conf/sigmod/Yuan0BF20}) &  \xmark& \checkmark &
                                                                     \checkmark &  \xmark\\
  
\textbf{ODT patterns ([our work])} & \checkmark &    \checkmark&\xmark &   \checkmark\\ 
 \hline
\end{tabular}
 \label{table:prevwork}
 \end{table}

\stitle{Outline} 
Section \ref{sec:relwork} reviews related work on spatio-temporal
pattern mining.
In Section \ref{sec:def}, we formally
define the problem we study in this paper. Section \ref{sec:algo}
presents an algorithm for extracting spatio-temporal flow patterns and
its optimizations.
In Section \ref{sec:ext}, we define interesting variants of
flow patterns and propose algorithms for their enumeration. In Section
\ref{sec:approximate2}, we propose an approximate method which
generates components of ODT patterns which participate in many atomic
patterns and tests combinations of them to identify potential patterns. 
Section \ref{sec:exps} evaluates our
methods on real networks with different characteristics. Finally,
Section \ref{sec:conc} concludes the paper with a discussion
about future work.

%
\section{Related Work}\label{sec:relwork}
The problem of enumerating, counting, and identifying spatiotemporal
patterns has been studied for over three decades, reflecting its
importance in data mining and in general in urban analytics \cite{DBLP:conf/sdm/GiannottiNP06,DBLP:journals/tits/AsifDGOFXDMJ14,DBLP:conf/kdd/Morimoto01,DBLP:conf/kdd/ZhangMCS04,DBLP:conf/icsdm/YooB11,DBLP:journals/eswa/Yu16,DBLP:journals/datamine/HanPYM04,DBLP:conf/edbt/KosyfakiMPT19,DBLP:journals/gis/CaiK22,DBLP:conf/icde/KosyfakiMPT21,DBLP:journals/air/AnsariAKBM20,DBLP:conf/vldb/HadjieleftheriouKBT05,zhang2018predicting,
DBLP:conf/wsdm/ParanjapeBL17,la2002spatio,tan2001finding,DBLP:conf/sdh/LaubeBK08,DBLP:journals/tgis/MennisL05,DBLP:conf/cikm/WangHL05,DBLP:conf/cikm/NanavatiCJK01,DBLP:journals/sensors/ZargariMM21,behara2022single,DBLP:journals/jits/BeharaBC22,DBLP:journals/tkdd/RongLDL24,DBLP:conf/kdd/FangLSX21,xia2021discovering,tu2022interday,pendyala2002time,djukic2013reliability,barcelo2022data,behara2021can,DBLP:conf/icdm/ChenNPTZB22,DBLP:journals/technometrics/XianYWL21,xian2021spatiotemporal,zou2021long,hazelton2001inference,chu2019deep,behara2021dbscan}. The main
goal of the problem is as follows: Given a graph $G$, identify spatial
events, correlations, or sequences that recur over time. Despite
considerable progress, existing methods still face challenges in
computational efficiency and scalability. The high computational cost
of current approaches has pushed researchers to explore alternative
problem formulations and faster algorithms. There is also a practical
need for methods that work with minimal input parameters (e.g., a
trajectory cannot be available due to privacy constraints) while
extracting useful patterns. Managing efficiency with pattern quality
under minimal user specification remains an open challenge.
In this section, we review works across several related areas:
sequential pattern mining, trajectory pattern mining, traffic flow
analysis, travel time estimations, and origin-destination flow
prediction. These areas provide context for our approach and highlight
current methods for discovering spatiotemporal patterns.

\stitle{Sequential and Trajectory Pattern Mining}
Sequential pattern mining methods
\cite{DBLP:conf/sdm/GiannottiNP06,DBLP:conf/icdm/CaoMC05} discover
interesting subsequences,
based on occurence frequency.
Agrawal and Srikant \cite{DBLP:conf/icde/AgrawalS95}
introduced the concept of sequential item-purchase patterns over a database of
time-ordered customer transactions.
Similar to sequence mining, trajectory pattern mining is the
problem of identifying frequent subsequences in space and time \cite{DBLP:conf/sigmod/NgC04,DBLP:reference/gis/GudmundssonLW08,
DBLP:journals/jips/KangY10, DBLP:conf/ssd/KalnisMB05,DBLP:conf/sigmod/TaoFPL04}. 
Spatio-temporal patterns are discovered from raw GPS data, which capture
routes or passenger movements.
Giannotti et al. \cite{DBLP:conf/kdd/GiannottiNPP07}
suggested the discovery of regions of
interest in trajectory pattern mining,
as considering sequences of predefined regions
reduces the complexity of the problem.
In the same spirit,
Cao et al. \cite{DBLP:conf/icdm/CaoMC05}
automatically segment the space into spatial regions by clustering the
tracked object locations; then, they identify
sequential patterns of regions that repeat themselves over time.
Choi et al., \cite{DBLP:journals/pvldb/ChoiPH17} introduce a tool
for discovering all regional movement patterns in semantic
trajectories.
In a semantic trajectory
\cite{DBLP:conf/gis/AlvaresBKMMV07}, each point of a trajectory not
only represents a specific location but also contains a semantic
category.
Patterns can then be defined based on semantics and locations over time.
Fan et al. \cite{DBLP:journals/pvldb/FanZWT16}
propose scalable trajectory mining methods using Spark.

Our problem is very different compared to previous work on
trajectory, sequence, and graph mining.
First, we are not interested in finding frequent paths
(subsequences, subgraphs), but in finding hot combinations of trip origins,
destinations, and timeslots. Second, we do not search for patterns at
the finest granularity only, but looking for patterns where any of
the three ODT components are generalized at an arbitrary granularity. Furthermore, we only use
a {\em weak monotonicity} property to construct and validate
generalized patterns from more detailed ones,
which means that the classic Apriori algorithm (and its variants) \cite{DBLP:conf/sigmod/AgrawalIS93,DBLP:journals/tkde/HanF99,DBLP:conf/ssd/KoperskiH95,DBLP:books/mit/fayyadPSU96/AgrawalMSTV96,DBLP:reference/gis/X08bq}
cannot be readily applied to solve our problem.

\stitle{Traffic Flow Mining} Mining traffic flow patterns \cite{DBLP:journals/tits/DuPWBWGLL20,DBLP:conf/kdd/FangLSX21,DBLP:journals/tkde/LiuLJXDTZ23,xie2020urban} is another related
problem
to ODT patterns.
Liu et al. \cite{DBLP:journals/tkde/LiuLJXDTZ23}  studied the
problem
of extracting traffic flow knowledge
from transportation data, i.e., pairs of POIs on the map that have
significant traffic flow between them.
Although we
also measure flow between regions, none of these previous
works studies flow patterns at {\em different granularities}, where regions
and time periods may consist of multiple atomic elements; hence,
previous works cannot be applied to solve our problem.
Moreover, they
require in-between locations, which may not be available and are not
relevant to our problem's objective.

\stitle{Travel Time Estimation} 
The main objective of \textit{Travel Time Estimation (TTE)} is to estimate the time to travel from a specific origin location on
the map to a specific destination at a given time
\cite{DBLP:journals/pvldb/00020B22,
  DBLP:journals/pvldb/JinLDW11, DBLP:conf/sigmod/Yuan0BF20}.
Lin et al., \cite{lin2023origin} propose an indexing approach (oracle)
for estimating the travel time between different regions
at different times of the day.
Although they also use the term $ODT$, our
goals are completely different. While our main objective is to
identify (generalized) Origin-Destination-Time patterns
considering the (passenger) flow, their goal is to predict the travel
time from a specific detailed region $O$ to a detailed destination $D$
at a time $T$.

\stitle{Origin-Destination Flow Prediction}
Predicting the origin-destination traffic flow has been a
well-studied problem in transportation research
\cite{ros2022practical,kumarage2023hybrid,DBLP:conf/bigdata2/YeZZXZY20,DBLP:journals/eswa/BeernaertsDLBW19,DBLP:journals/eaai/PamulaZ23,DBLP:conf/icde/ShiYGLZYLL20,DBLP:journals/eswa/YuZGZSZZY25,DBLP:conf/ijcnn/YungYLT24,duan2019prediction,DBLP:conf/kdd/WangYCW0019,DBLP:journals/jits/BeharaBC22,behara2021dbscan,DBLP:journals/tkde/GongLZLZ22}. The
main objective is to estimate the flow (e.g., number of
passengers in a transportation network or cars in a road network) between
origin-destination pairs.
A typical approach is to construct a
Origin-\linebreak Destination matrix \cite{ashok1996estimation}, where each cell
represents the trips between an
origin and a destination. Tu et al., \cite{tu2022interday} study the
stability of travel patterns on different times of the day and use
predefined geographic units to group the origin-destination pairs of
each journey. Behara et al., \cite{DBLP:journals/jits/BeharaBC22}
study the problem of OD flow prediction by proposing a
geographical window-based structural similarity
(GSSI) index that captures statistics of geographically
correlated OD pairs.
OD pairs correlate if they share similar structural
properties and therefore expand to neighbor regions.
Their proposed solution can help identifying local
patterns more efficiently.
Similarly, \cite{behara2021dbscan} proposes a DBSCAN framework
to identify travel patterns in multi-density high dimensional matrices.
Although
\cite{DBLP:journals/jits/BeharaBC22,behara2021dbscan} share similarities with our
work, the set up of the problem is different.
First, they use similarity
metrics to group
regions together when defining OD pairs.
In our case, we perform region and time generalization based on
(geographic) adjacency and connectivity in a neighborhood graph.
Wang et al.,
\cite{DBLP:conf/kdd/WangYCW0019} develop a model for predicting the
flow density in different origin-destination pairs. To this end, they represent a city as
a grid and take into consideration regions that may have a significant
amount of flow compared to other less important regions. The goal of
this work is similar to ours, however they use additional data such as
POIs and area
proximity, which, in our case, are not available. In summary,
our framework can identify flow patterns that can be used to predict
demands in the flow between generalized regions and at (generalized)
timeslots 
efficiently using the minimum available information.
\section{Definitions}\label{sec:def}
In this section, we formally define ODT patterns and the graph
wherein they are identified.
Table \ref{tab:notations} shows the
notations used frequently in the paper.

\begin{table}[ht]
  \caption{Table of notations}\label{tab:notations}
\centering
\footnotesize
\begin{tabular}{|c|c|}
\hline
Notations & Description \\
\hline  
$G(V,E)$ & region neighborhood graph  \\
$r_i$ & atomic region \\
$R_i$ & region \\
$t_i$ & atomic timeslot \\
$T_i$ & timeslot\\
$P$ & atomic ODT pattern or triple \\
$P.O$ & pattern/triple origin \\
$D$ &  pattern/triple destination \\
$T$ &  pattern/triple timeslot \\
$\sigma(P)$ & support of atomic ODT pattern $P$ \\
$s_a$ & support of atomic regions and timeslots \\
$s_r$ & ratio threshold \\   
$P$.cnt & number of atomic patterns in ODT pattern $P$ \\
$\mathcal{P_\ell}$/$\mathcal{T_\ell}$ & Set of ODT patterns/triples at
  level $\ell$\\
\hline         
\end{tabular}
\end{table}

The main input to our problem is a {\em trips} table,
which
records information about trips from origins to destinations at
different times. Each origin/destination is a 
minimal region of interest on a map (e.g., a
district, a metro station, etc.), called {\em atomic region}.
Atomic regions are specified by the application; they could be
pre-defined districts or stations/stops and their nearby area.
Spatial clustering can also be applied on precise locations (e.g.,
taxi pickup/dropoff locations) to define a set of atomic regions, as
in \cite{DBLP:conf/icdm/CaoMC05}.
Let $V$ be the set of all atomic regions.
An undirected graph
$G(V,E)$ defines the neighboring relations between atomic regions;
there is an edge
$(v,u)$ in $E$ iff $v\in V$ and $u\in V$ are neighbors on the
map.
Finally, the timeline is divided into periods that repeat themselves (e.g., 24-hours each) 
and each period is discretized into time ranges (e.g., 48 30-minute
slots).
Discretization of time can be uniform or it may follow the data distribution
(e.g., using the boundaries of an equi-depth histogram).
Each minimal time range of interest is called {\em atomic} timeslot.
Periods can also be classified (e.g., weekdays vs. weekends); hence
atomic timeslots may refer to different period classes (e.g.,
8:00-8:30 on weekdays).



  \begin{definition} [Region/Timeslot] \label{def:eregion}
 A  region $r$ is a subset $V'$ of $V$, such that the
induced subgraph $G'(V',E')$ of $G$ is connected. A
timeslot $T$ is a continuous sequence of atomic timeslots.
  \end{definition}



\begin{definition} [Generalization of a region/timeslot] \label{def:genrt}
 A region $R_1$ is a generalization of region $R_2$ iff
 $R_2\subset R_1$. A timeslot $T_1$ is a 
generalization
of timeslot $T_2$ iff
$T_2\subset T_1$.
\end{definition}

\begin{definition} [Minimal generalization of a region/timeslot] \label{def:mingenrt}
 A region $R_1$ is a minimal generalization of region $R_2$ iff
 $R_2\subset R_1$
and  $R_1 - R_2$ is an atomic region. A timeslot $T_1$ is a minimal
generalization
of timeslot $T_2$ iff
$T_2\subset T_1$ and $T_1 - T_2$ is an atomic timeslot.
\end{definition}

For example, region $\{B,C,D\}$ is a minimal generalization of $\{B,D\}$.
Symmetrically,
$\{B,D\}$ is a  minimal specialization of $\{B,C,D\}$.

We found that the start and end times of individual trips between the same origin
and destination are strongly correlated.
Hence, we can map each trip in the trips table
to an {\em atomic ODT triple} $(o,d,t)$, where
$o$ is the origin region of the trip (GPS locations can be mapped
to the nearest spatially $v\in V$), $d$ is the trip's destination
region,
and $t$ is the atomic timeslot that contains the trip's
origin time.%
\footnote{Our definitions and techniques can be extended for data inputs
where there is no correlations between the begin and end times of
trips; in this case, we should map trips to $(o,d,st,et)$ quadruples
and patterns should be ODSE quadruples.}

\begin{definition} [Atomic ODT triple] \label{def:atriple}
  A triple $(o,d,t)$ is atomic if:
  \begin{itemize}
  \item $o$ is an atomic region
  \item $d$ is an atomic region
  \item $t$ is an atomic timeslot
  \item $o \neq d$
  \end{itemize}
\end{definition}

\begin{figure}[h]
      \centering
      \includegraphics[width=0.85\columnwidth]{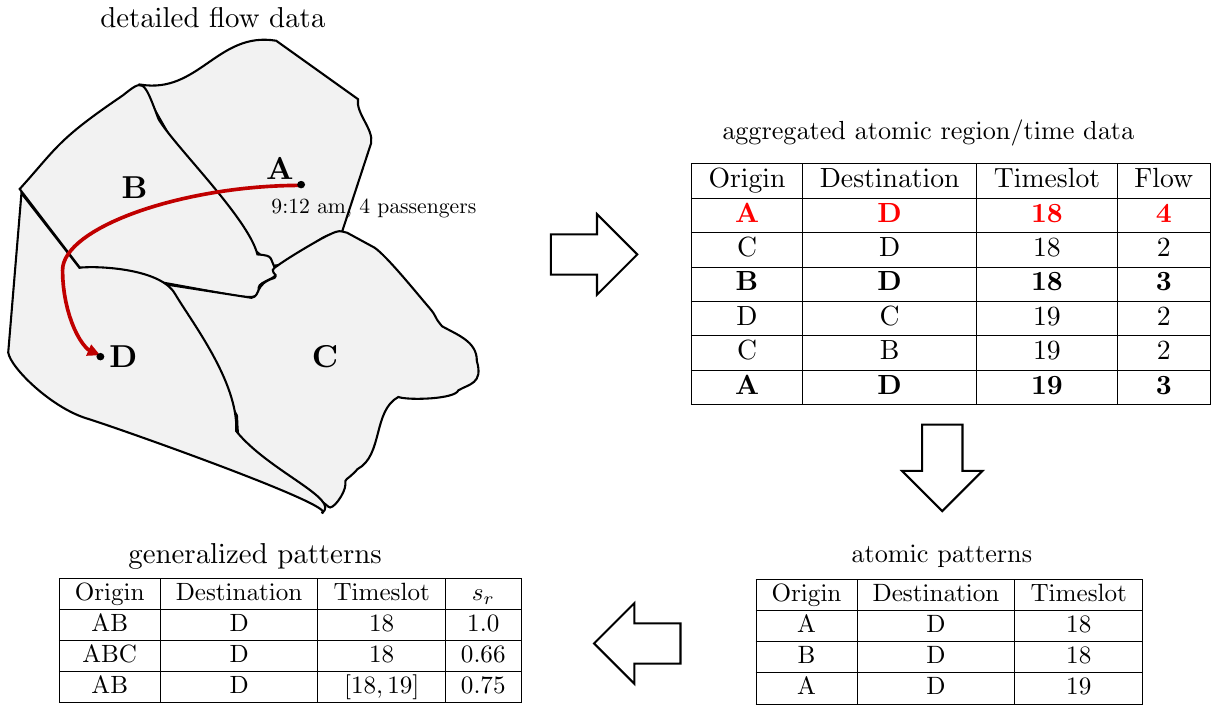}
      \caption{A detailed example}
      \label{fig:detailed}
\end{figure}

Given an atomic ODT triple $P$, the
{\em support} $\sigma(P)$ of $P$ is the total number of passengers
(flow) of the trips that are mapped to $P$.
For example, the top-left of Figure \ref{fig:detailed} shows a map and
an individual trip in it. The top-right of Figure \ref{fig:detailed} has
the {\em aggregated trips table}, which contains all atomic $(o,d,t)$
triples, after aggregating all trips that correspond to the same
$(o,d,t)$. For instance, trips ($B$, $D$, 9:20, 2) and ($B$, $D$,
9:29, 1) are merged to triple $(B,D,18)$ with total flow $3$.%
\footnote{All time moments between
  9:00 and 9:30 are generalized to timeslot 18, which is the 18th slot
  in 30-minute intervals, starting from 00:00-00:30 (mapped to 0).}
Next, we define generalized (i.e., non-atomic) ODT
triples.




\begin{definition} [ODT triple] \label{def:etriple}
An ODT triple $(O,D,T)$ consists of a region $O$, a region $D$, and a
timeslot $T$, such that $O\cap D = \emptyset$.
\end{definition}

\begin{definition} [ODT triple generalization] \label{def:gpattern}
  An ODT triple $P_1$ is a generalization of ODT triple $P_2$ if
for each $X\in \{O,D,T\}$, either $P_1.X=P_2.X$ or $P_2.X\subset
P_1.X$,
and for at least one $X\in \{O,D,T\}$, $P_2.X\subset
P_1.X$.
\end{definition}

\begin{definition} [Minimal generalization of ODT triple] \label{def:mingen}
  An ODT triple
  $P_1$ is a minimal generalization of ODT triple $P_2$ if one of the
  following holds:
    \begin{itemize}
  \item $P_1.O=P_2.O$, $P_1.D=P_2.D$ and $P_1.T$ is a minimal generalization of $P_2.T$  \item $P_1.D=P_2.D$, $P_1.T=P_2.T$ and $P_1.O$ is a minimal generalization of $P_2.O$  \item $P_1.O=P_2.O$, $P_1.T=P_2.T$ and $P_1.D$ is a minimal generalization of $P_2.D$ \end{itemize}
\end{definition}


We now turn to the definition of ODT patterns; we start by atomic ODT
patterns and then move to the generalized  ODT patterns.

\begin{definition} [Atomic ODT pattern] \label{def:atomicpattern}
Let $AT_r$ be the set of atomic ODT triples with non-zero support.
Given a threshold $s_a, 0< s_a\le 1$, an
atomic ODT triple $P$ is called an atomic ODT pattern if $\sigma(P)$ is
in the top $s_a\times |AT_r|$ supports of triples in $AT_r$.
\end{definition}

The rationale of using $s_a$ is that we want to restrict the set of
important  atomic ODT triples, by considering as patterns the triples
that have large total flow (accumulated over the repeated instances of
the triples in the trips table).
Figure \ref{fig:detailed} (bottom-right) shows the atomic ODT patterns for our running example if $s_a=0.5$.

\begin{definition} [ODT pattern] \label{def:pattern}
An ODT pattern $P$ is an ODT triple where:
\begin{itemize}
\item the ratio of atomic triples in $P$, which are atomic patterns, is at
least equal to a {\em minimum ratio} threshold $s_r$
\item there exists a minimal specialization of $P$, which is an ODT pattern  
\end{itemize}  
\end{definition}

The number of atomic triples in $P$, which are atomic patterns is
denoted by  $P$.cnt.
In the example of Figure \ref{fig:detailed}, 
if $s_r=0.6$, 
$(AB,D,18)$ is a (generalized) ODT pattern where origins $A$ and $B$
have significant joint flow to destination $D$ at timeslot $t=18$,
because the pattern includes two out of three atomic patterns. The
second condition of Definition \ref{def:pattern} is a {\em sanity} constraint,
which prevents a generalized triple $P$ from being characterized as a pattern
if there is no minimal specialization of $P$ that is also a pattern;
intuitively, a pattern should have at least one minimal specialization which
is also a pattern (weak monotonicity).

A pattern (triple) $P$ is said to be level-$\ell$ pattern (triple) if the total number of
atomic elements in it (regions and timeslots) is $\ell$. Hence, atomic
patterns are level-3 patterns, since they contain exactly 3 elements
(i.e., two atomic regions and one atomic timeslot). Similarly, triple
$(A, BC, [1,3])$ is a level-6 triple because it includes 1 atomic
region in its origin, 2 atomic regions in its destination, and 3
atomic timeslots in its time-range (note that $[1,3]$ includes atomic
timeslots $\{1,2,3\}$).


\section{Pattern extraction}\label{sec:algo}
To find the ODT patterns, 
we first start by finding the atomic ODT patterns, i.e., the $(o,d,t)$
triples which are frequent/significant, where $o$ and $d$ are atomic
regions and $t$ is an atomic timeslot.
This is trivial and can be done by one pass over the aggregated trips data, where the
occurrence of each $(o,d,t)$ triple is unique and by selecting the top
$s_a$ ratio of them as atomic patterns. 
Our most challenging task is then to define an algorithm that
progressively synthesizes non-atomic patterns from atomic patterns and
prunes the search space effectively.

Recall that a non-atomic triple $P=(O,D,T)$ is a pattern
if at least a ratio $s_r>0$ of its included atomic triples
are patterns. Hence, by definition, a non-atomic pattern generalizes at
least one atomic pattern $(o,d,t)$.
The
pattern synthesis algorithm uses the set of atomic patterns
and the
region neighborhood graph $G$ to form the non-atomic patterns.
Given an existing $(O,D,T)$ pattern $P$, we 
attempt a minimal generalization of $P$ by including into the
set $O$ a neighboring atomic region to the existing regions in $O$, or
doing the same for set $D$, or adding an atomic neighboring timeslot
to $T$.

The challenge is to
prune candidate generalizations that cannot be patterns.
For this, we need a fast way to compute (or bound) the number of
contributing (newly added to $P$) atomic patterns to the ratio of the
candidate.



\subsection{Baseline Algorithm}\label{sec:baseline}
We now present a baseline algorithm for enumerating all the atomic and
extended ODT patterns in an input graph $G(V,E)$.
The first step of Algorithm \ref{algo:baseline} is to scan all trips
data and compute the support counts of all atomic triples $\mathcal{T}_3$.
Then, it finds the set $\mathcal{P}_3$ of atomic patterns, i.e., the triples
having support count at least equal to
$minsup$, which is the support
count of the $s_a\cdot|\mathcal{T}_3|$-th triple in $\mathcal{T}_3$ in
decreasing support order.
All triples (patterns) in $\mathcal{T}_3$ ($\mathcal{P}_3$) have exactly
three atomic elements (regions or timeslots).
The algorithm progressively finds the patterns with more atomic
elements.
Recall that a triple (pattern) $P$ is at level $\ell$, i.e., in set $\mathcal{T}_\ell$
($\mathcal{P}_\ell$) if it has $\ell$ atomic elements; we also call $P$ a
level-$\ell$ triple (pattern).
Candidate patterns $CandP$ at level $\ell+1$ are generated by either adding an atomic
region at $O$ or an atomic region at $D$ or an atomic timeslot at $T$,
provided that the resulting triple is valid according to Definition \ref{def:etriple}.
If a $CandP$ has been considered before, it is disregarded. This may
happen because the same triple can be generated from two or more
different triples at level $\ell$. For example, candidate pattern
$(AB,C,1)$ could be generated by pattern $(A,C,1)$ (by extending
region $A$ to region $AB$) and by $(B,C,1)$ (by extending
region $B$ to region $AB$).
Hence, we keep track at each level $\ell$ the set of triples that
have been considered before, in order to avoid counting the same
candidate twice.%
\footnote{Since a pattern at level $\ell+1$ requires at least one
  and not all its minimal specializations to be patterns, an id-numbering
  scheme for atomic regions, which would extend patterns by only
  adding elements that have larger id would not work. For example, if
  both $(A,C,1)$ and  $(B,C,1)$ are patterns, $(AB,C,1)$ can be
  generated by both of them; however, if just $(B,C,1)$ is a pattern,
  $(AB,C,1)$ can only be generated by  $(B,C,1)$.}

To check whether a candidate $CandP$ not considered before is a
pattern,
we need to divide the number $CandP$.cnt of atomic patterns included in $CandP$
by the total  number $CandP$.card of atomic triples in $CandP$.
If this ratio is at least $s_r$, then $CandP$ is a pattern.
$CandP$.card can be computed algebraically: it is the product of
atomic number elements in each of the three ODT components. For example,
$(AB,CD,[1,3])$.card = $2\cdot 2\cdot 3=12$ because there are 12
atomic triples in $(AB,CD,[1,3])$, i.e., combinations of elements
$\{A,B\}$,  $\{C,D\}$, and $\{1,2,3\}$.
To compute $CandP$.cnt fast, we can take advantage of the fact that we
already have $P$.cnt, i.e., the number of atomic patterns in the
generator pattern. We only have to compute the $P'$.cnt for the
{\em difference} $P'=CandP-P$ between $CandP$ and $P$, which is the
triple consisting of the extension element in the extended dimension
(one of O, D, T) together with the element-sets in the intact
dimensions (two of O, D, T).
For example, if $P=(A,CD,[1,2])$ and $CandP=(AB,CD,[1,2])$, then
$P'=(B,CD,[1,2])$.
To compute $P'$.cnt,
Algorithm \ref{algo:baseline} enumerates all atomic triples in $P'$ to
check whether they are patterns.
It then sums up $P$.cnt and $P'$.cnt to derive $CandP$.cnt.


\begin{algorithm}
\begin{algorithmic} [1]
 \scriptsize
\Require a region graph $G(V,E)$; a trips table;
a minimum support $s_a$ for atomic ODT patterns; a minimum support
ratio $s_r$ for
non-atomic ODT patterns 
\State $\mathcal{T}_3$ = atomic triples computed from trips table
\State $\mathcal{P}_3$ = triples in $\mathcal{T}_3$ with support $\geq
s_a\times |\mathcal{T}_3|$
\For{all atomic triples $P\in \mathcal{T}_3$}
       \State $P$.cnt = 1 if $P\in \mathcal{P}_3$, else $P$.cnt=0
 \EndFor   
\State $\ell$ = 3
\While {$|\mathcal{P_\ell}| > 0$}  \Comment{$\mathcal{P}_\ell$ = set of
  level-$\ell$ patterns}          
     \State $\mathcal{P}_{\ell+1}$ = $\emptyset$   \Comment{Initialize pattern set at level
       $\ell+1$}
     \For{each $P$ in $\mathcal{P}_\ell$}
         \For{each minimal generalization $CandP$ of $P$}
               \If{$CandP$ not considered before}
                      \State $P'$= $CandP-P$ 
                      \State $CandP$.cnt = $P$.cnt + $P'$.cnt
                     \If{$CandP$.cnt $/$ $CandP$.card $\geq s_r$}
                           \State add $CandP$ to $\mathcal{P}_{\ell+1}$
                       \EndIf
                 \EndIf
            \EndFor
     \EndFor
     \State $\ell$ = $\ell$ + 1       
\EndWhile                                 
\end{algorithmic}
\caption{Baseline Algorithm for finding all ODT patterns}
\label{algo:baseline}
\end{algorithm}

\begin{figure}[h]
      \centering
      \includegraphics[width=0.85\columnwidth]{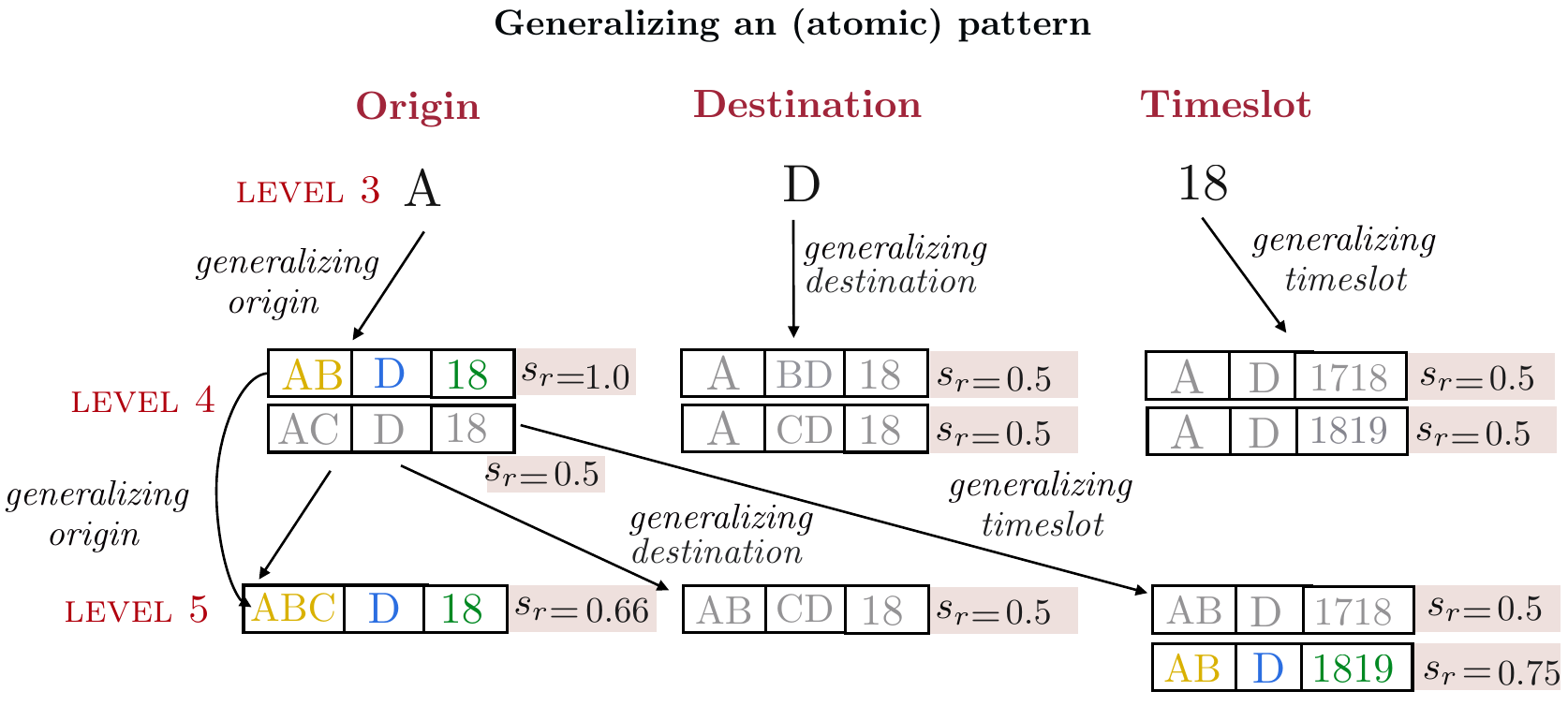}
      \caption{Pattern enumeration example}
      \label{fig:lattice}
\end{figure}

Figure \ref{fig:lattice} exemplifies the pattern enumeration
process in our running example (see Figure \ref{fig:detailed}).
Atomic pattern $P=(A,D,18)$ can be generalized by adding to the origin
any of the neighbors of atomic region $A$, to the destination any of the neighbors
of atomic region $D$, and to timeslot 18 either timeslot 17 or
timeslot 19. Each of these generalization forms a candidate pattern
$CandP$ at level 4. Counting the support of these candidates requires
counting only the difference $P'$. For example, to count the support
of $(AB,D,18)$, we only have to add to the support of $P=(A,D,18)$ the
support of $P'=(B,D,18)$, which is 1. Then, the support of $(AB,D,18)$
is found to be 2. Assuming that $s_r=0.6$, $CandP=(AB,D,18)$ is a pattern,
since the ratio of atomic patterns in it is $1.0\ge s_r$. All patterns
that stem from $P=(A,D,18)$ up to level 5 are emphasized in Figure
\ref{fig:lattice}; these are used to generate candidate patterns at
the next levels. 

\stitle{Complexity Analysis}
In the worst case, all valid combinations of regions and timeslots can
be considered as candidate patterns.
In other words, each subset of $V$ can be the union of O and D and for
each such subset of size $k$ can be split in $2^k-2$ ways.
Hence, the number of possible OD pairs is 
$\sum_{k=1}^{|V|-1}{|V| \choose k}(2^k-2)$.
If $S$ is the number of atomic timeslots, then the number of
possible (generalized) timeslots to be included in a candidate pattern
is $2^{|S|}$. Hence, the worst-case space/time complexity of ODT pattern
enumeration is $O\left
  (2^{|S|}\sum_{k=1}^{|V|-1}{|V| \choose k}(2^k-2)\right )$.

\subsection{Optimizations}\label{sec:opt}
We now discuss some optimizations to the baseline algorithm, which can
greatly enhance its performance in practice.

{\bf Avoid re-counting $P'$.}
The first approach is based on {\em caching} the ODT triples that have
been counted before. Instead of computing $P'$.cnt directly for
$P'=CandP-P$, we first check whether $P'$.cnt is already
available. This requires us to cache the counted triples and their
supports at each level in a hash table. Hence, before counting $P'$,
we first search the hash table, which caches the triples at level
$|P'|$ to see if $P'$ is in there. In this case, we simply use
$P'$.cnt instead of computing it again from scratch.

{\bf Fast check for zero support of $P'$.}
The second optimization is based on the observation that for some
pairs $(o,d)$ of atomic regions,
there does not exist any timeslot $t$, such that
$(o,d,t)$ is an atomic pattern in $\mathcal{P}_3$. For example, if $o$ and $d$
are remote regions on the map, it is unlikely that there is
significant passenger flow that connects them at any time of the day.
We take advantage of this to avoid counting any $P'$ which may not
include atomic patterns. Specifically, for each atomic region $r$, we
record (i) $r.dests$, the set of atomic regions $r'$, such that there
exists a $(r,r',t)$ pattern in $\mathcal{P}_3$; and (ii)
$r.srcs$, the set of atomic regions $r'$, such that there
exists a $(r',r,t)$ pattern in $\mathcal{P}_3$.
If, in Algorithm $\ref{algo:baseline}$, $CandP$ is a minimal
generalization of $P$, by expanding $P.O$ to include a new atomic
region $r$, then $P'=CandP-P$ should only include $r$ in $P'.O$. If
$P'.D\cap r.dests = \emptyset$, then $P'$ does not include any atomic
patterns and $CandP$.cnt is guaranteed to be equal to $P$.cnt. Hence,
we can skip support computations for $P'$.
Symmetrically, if $CandP$ is a minimal
generalization of $P$, by expanding $P.D$ to include a new atomic
region $r$, then $P'=CandP-P$ should only include $r$ in $P'.D$. If
$P'.O\cap r.srcs = \emptyset$, then $CandP$.cnt is guaranteed to
be equal to $P$.cnt.
Overall, by keeping track of $r.dests$ and $r.srcs$ for each atomic
region $r$, we can save computations when counting the supports of patterns. 
Since the space required to store $r.dests$ and $r.srcs$ is $O(|V|)$
in the worst case, the total space complexity of these sets is
$O(|V|^2)$. This cost is bearable, because our problem typically
applies on transportation networks or district neighborhood graphs in
urban maps,  where the number of vertices in $V$ is rarely large.  


{\bf Improved neighborhood computation.}
The minimal generalizations of a pattern $P$ are generated by
minimally generalizing $P.O$, $P.D$, and $P.T$.
The generalization of $P.T$ is trivial as we add one atomic
timeslot before the smallest one in  $P.T$ or after the largest one in
$P.T$.
On the other hand, computing the minimal generalizations of a region
$R$ (i.e., $P.O$ or $P.D$) can be costly if done in a brute-force
way. The naive algorithm tries to add to $P.O$ all possible neighbors of each
atomic region $r\in R$ and for each such neighbor not in $P.O$ and
$P.D$ it measures the support of the corresponding generalized
pattern $P'$, if $P'$ was not considered before. Since the same $P'$
can be generated by multiple $P$,  checking whether  $P'$ has been
considered before can be performed a very large number of times with a
negative effect in the runtime. We design a neighborhood computation
technique for a region $R$, which avoids generating the same $P'$
multiple times. The main idea is to collect first all neighbors of all
$r\in R$ in a set $N$ and then compute (in one step) $N-P.O-P.D$,
i.e., the set of regions $r$ that minimally expand $R$ to form the
minimal generalizations of $P$. 

\begin{figure}[h]
      \vspace{-0.2in}
      \centering
      \includegraphics[width=0.75\columnwidth]{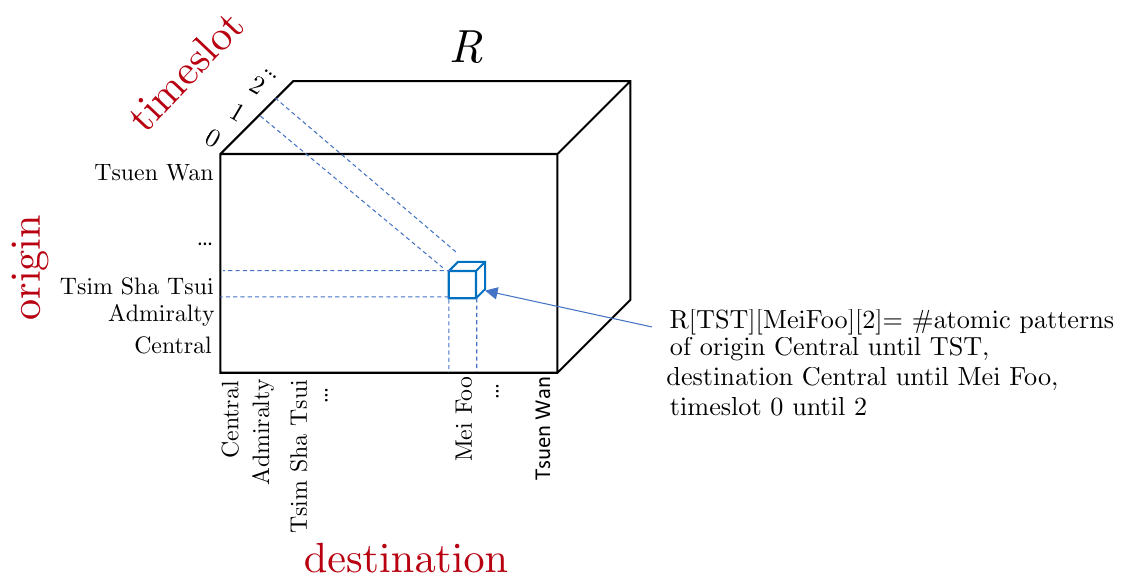}
      \caption{Prefix sum example}
       \vspace{-0.45cm}
      \label{fig:prefix}
\end{figure}

{\bf Indexing atomic patterns.}
As another optimization, we employ a prefix-sum index which can help
us to compute an upper bound of the support of $P'$.
The main idea comes from indexes used to compute range-sums in OLAP
\cite{DBLP:conf/sigmod/HoAMS97}.
Let $N$ be the number of atomic regions and $M$ be the number of
atomic timeslots.
Consider a $N\times N \times M$
array $A$, where each cell corresponds to an atomic ODT
triple.
The cell includes a 1 if the corresponding   atomic ODT
triple is a pattern; otherwise the cell includes a  0.
In addition, consider a 3D array $R$ with shape $(N+1)\times (N+1) \times (M+1)$.
Each element $R[i][j][k]$ of $R$ is the sum of all elements
$A[i'][j'][k']$ of $A$, such that $i'\le i$, $j'\le j$, and $k'\le k$
$R[i][j][k]=0$ if any of $i,j,k$ is 0. $R$ is the {\em
  prefix-sum} array of $A$, as illustrated in 
Figure \ref{fig:prefix}.

Consider a 3D range $[a,b], [c,d], [e,f]$, where $0<a\le b \le N$,
$0<c\le d \le N$, and $0<e\le f \le M$.
The sum of values in $A$
inside this range can be computed as
follows:
\begin{align*}
&R[b][d][f]\\
&-R[a-1][d][f]-R[b][c-1][f]-R[b][d][e-1]\\
&+R[a-1][c-1][f]+R[b][c-1][e-1]+R[a-1][d][e-1]\\
&-R[a-1][c-1][e-1]
\end{align*}


Now, consider a $P'$ which needs to be counted. $P'$ includes a set of
atomic origin regions, a set of atomic destination regions, and a set
of atomic timeslots.
The atomic timeslots are guaranteed to be a continuous sublist of
regions in the corresponding dimension of the 3D array $A$, starting,
say, from timeslot $e$ and including up to timeslot $f$.
However, the region sets in $P'$ are not guaranteed to be continuous.
Still, the 3D range $[a,b], [c,d], [e,f]$, where $a$ ($c$) is the
origin (destination) region in $P'$ with the smallest ID and 
$b$ ($d$) is the
origin (destination) region in $P'$ with the largest ID is
guaranteed to be a superset of atomic
triples in $P'$.
Hence, the prefix-sum index can give us in $O(1)$ time an {\em upper
  bound} of the number of atomic patterns in $P'$.
If this upper bound is added to the support of $P$ and the resulting
support is less than $s_r$, then $CandP$ is definitely not a pattern,
so we can avoid counting $P'$.
To maximize the effectiveness of this optimization, we
give IDs to regions
by applying a depth-first traversal of $G$ from the vertex
with the smallest degree.

\section{Pattern Variants}\label{sec:ext}
In this section, we explore alternative problem definitions and the
corresponding ODT pattern extraction algorithms that can be more useful than our
general definition in certain problem instances.
In particular, we observe that the number of patterns can be huge even
if relatively small $s_a$ and large
$s_r$ are used.
In addition, setting global
thresholds may not be ``fair'' for some regions which are
under-represented in the data. 
To address these issues, we propose (i) size-bounded patterns, (ii)
constrained-pattern search,
and (iii) rank-based patterns.

\subsection{Size-bounded Patterns}\label{sec:sizebounded}
The first type of constraint that we can put to limit the number of
patterns is on the sizes of the pattern O,D, and T components.
Specifically, we can set an upper bound $B_O$ to the
number of atomic regions in an origin region of a pattern. Similarly,
we can limit the number of regions in $D$ to at most $B_D$ and the
number of atomic timeslots to at most $B_T$. In effect, this limits
the number of levels that we use for pattern search to $B_O\cdot
B_D\cdot B_T$.
For pattern enumeration, we use the same algorithms and optimizations
discussed in Section \ref{sec:algo}, but with the constraints applied
whenever we expand a pattern to generate the candidate patterns at the
next level.

\subsection{Constrained Patterns}\label{sec:restricted}

Another way to control the number of the patterns, but also focus on
specific regions and/or timeslots that are under-represented in the
entire population is to limit the domain of atomic regions and
timeslots.
Specifically, we give as parameter to the problem the set of
atomic regions $V_O\subseteq V$ that we are interested in to act as
origins the set $V_D\subseteq V$ of regions that can act as
destinations, and $T_R\subseteq T$, a restricted contiguous
subsequence of the entire sequence of atomic timeslots $T$ to be used as
timeslots in the patterns.
The induced subgraphs by $V_O$ and $V_D$ should be connected, in order
to potentially  have the entire $V_O$ (and/or  $V_D$) as an origin
(destination) of a pattern.
For example, if a data analyst is interested in flow patterns from South
Manhattan to Queens in afternoon hours, she could include in $V_O$
(resp. $V_D$) all the atomic regions in South
Manhattan (resp. Queens) and restrict the timeslots to be used in
patterns to only afternoon hours.
Depending on the sizes of $V_O$, $V_D$, and $T_R$ pattern enumeration
can be significantly faster compared to unconstrained pattern search.

\subsection{Rank-based patterns}\label{sec:rank}
Another way to control the number of patterns and still not miss the most
important ones is to regard as patterns, at each level,
the $k$ triples with the highest support 
and prune the rest of them as non-patterns.
This is achieved by replacing the minimum ratio threshold $s_r$ by
a parameter $k$, which models the ratio of eligible triples
at each level which are considered to be important.
More formally, let $\mathcal{T}_\ell$ be the set of triples at level
$\ell$, which are minimal generalization of patterns at level $\ell-1$.
The patterns $\mathcal{P}_\ell$ at level $\ell$ are the 
$k$ triples in $T_{\ell}$ having the largest number of atomic patterns.

\begin{definition} [ODT pattern (rank-based)] \label{def:rankpattern}
An ODT triple $P$ at level $\ell$ is a rank-based ODT pattern if:
\begin{itemize}
\item there exists a minimal specialization of $P$ which is a
  rank-based ODT pattern  
\item there are no more than $k$ minimal generalizations
  of level-$(\ell-1)$ rank-based ODT patterns that include more frequent
atomic patterns than $P$. 
\end{itemize}  
\end{definition}





\subsubsection{Baseline approach for rank-based pattern enumeration}
\label{sec:baselinerank}
A baseline approach for enumerating rank-based patterns is to
generate all eligible triples at each level $\ell$, which are minimal
generalizations of patterns at level $\ell-1$. For each such triple,
count its support (i.e., number of atomic patterns included in
it). We may use the optimizations proposed in Section \ref{sec:opt},
to reduce the cost of generating ODT triples that are candidate
patterns and counting their supports.
After generating all triples and counting their
supports, we select the top-$k$ ones as patterns.
Only these patterns are used to generate the candidate patterns at
level $\ell +1$.

\subsubsection{Optimized rank-based pattern enumeration}
\label{sec:optrank}
To minimize the number of generated triples at each level $\ell$ and
the effort for counting them, we
examine the patterns at $\ell-1$ in decreasing order of their
potential to generate triples that will end up in the top-$k$
triples at level $\ell$.
Hence, we access the patterns $P$ at level $\ell-1$  in decreasing order
of their support $P$.cnt. For each such pattern $P$ and for each minimal
generalization $CandP$ of $P$, we first compute the potential of
$P'=CandP-P$ to add to the support $CandP$.cnt (initially $CandP$.cnt
= $P$.cnt). If, by adding the maximum possible $P'$.cnt to
$CandP$.cnt, $CandP$.cnt cannot make it to the top-$k$ $\ell$-triples
so far, then we prune $CandP$ and avoid its counting.
The maximum possible $P'$.cnt can be computed based on the following
lemma:

\begin{lemma}\label{lemma:maxsupport}
The maximum possible $P'$.cnt that can be added to $P$.cnt, to derive
the support of $CandP$ is as follows:
\begin{itemize}
  \item If $CandP$ is generated by minimally generalizing $P.O$, then
    $P'$.cnt equals $|P.D|\cdot |P.T|$.    
  \item If $CandP$ is generated by minimally generalizing $P.D$, then
    $P'$.cnt equals $|P.O|\cdot |P.T|$.
  \item If $CandP$ is generated by minimally generalizing $P.T$, then
    $P'$.cnt equals $|P.O|\cdot |P.D|$.   
\end{itemize} 
\end{lemma}

Let $\theta$ be the $k$-th largest support of the 
triples generated so far at level $\ell$.
If for the next examined $P$ from level $\ell-1$ to generalize,
$P$ cannot be generalized to a $CandP$ that may end up in the top-$k$
level-$\ell$ triples, then we can immediately prune $P$. The condition
for pruning $P$ follows:

\begin{lemma}\label{lem:prunep}
If $P$.cnt + $\max\{|P.D|\cdot |P.T|, |P.O|\cdot |P.O|, |P.O|\cdot
|P.D|\}$ $\le\theta$, then no minimal generalization of $P$ can enter
the set of top-$k$ level-$\ell$ ODT triples.
\end{lemma}

\begin{algorithm}
\begin{algorithmic} [1]
 \scriptsize
\Require a region graph $G(V,E)$; a trips table;
a minimum support $s_a$ for atomic ODT patterns; number $k$ of top
patterns to be generated at each level; maximum level considered ($maxl$) 
\State $\mathcal{T}_3$ = atomic triples computed from trips table
\State $\mathcal{P}_3$ = triples in $\mathcal{T}_3$ with support $\geq s_a$
\For{all atomic triples $P\in \mathcal{T}_3$}
       \State $P$.cnt = 1 if $P\in \mathcal{P}_3$, else $P$.cnt=0
 \EndFor   
\State $\ell$ = 3
\While {$|\mathcal{P_\ell}| > 0$ and $\ell< maxl$}  \Comment{extend
  level-$\ell$ patterns}          
     \State $\mathcal{P}_{\ell+1}$ = $\emptyset$   \Comment{Initialize
       $k$-minheap with level-$(\ell+1)$ patterns}
     \For{each $P$ in $\mathcal{P}_\ell$ in decreasing order of
       $P$.cnt}
         \If {$|\mathcal{P}_{\ell+1}|=k$ and $P$.cnt+$\max\{|P.D|\cdot |P.T|, |P.O|\cdot |P.O|, |P.O|\cdot
           |P.D|\}\le \mathcal{P}_{\ell+1}$.top.cnt}
         \State {\bf continue} \Comment{Prune $P$ based on Lemma \ref{lem:prunep}}
         \EndIf
         \If {$|\mathcal{P}_{\ell+1}|=k$ and $P$.cnt+ $|P.D|\cdot
           |P.T|\le \mathcal{P}_{\ell+1}$.top.cnt}
            \For{each minimal generalization $CandP$ of $P$ by origin}
               \If{$CandP$ not considered before} \label{lin:st}
                  \State $P'$= $CandP-P$
                  \State$CandP$.cnt = $P$.cnt + $P'$.cnt
                  \If{$|\mathcal{P}_{\ell+1}|<k$}
                     \State add $CandP$ to $\mathcal{P}_{\ell+1}$
                  \Else
                     \If{$CandP$.cnt $>\mathcal{P}_{\ell+1}$.top.cnt}
                     \State update $\mathcal{P}_{\ell+1}$ with $CandP$
                     \EndIf
                  \EndIf
               \EndIf \label{lin:end}
            \EndFor
          \EndIf
          \If {$|\mathcal{P}_{\ell+1}|=k$ and $P$.cnt+ $|P.O|\cdot
           |P.T|\le \mathcal{P}_{\ell+1}$.top.cnt}
            \For{each minimal generalization $CandP$ of $P$ by
              dest.}
            \State Lines \ref{lin:st} to \ref{lin:end} above 
            \EndFor
            \EndIf
          \If {$|\mathcal{P}_{\ell+1}|=k$ and $P$.cnt+ $|P.O|\cdot
           |P.D|\le \mathcal{P}_{\ell+1}$.top.cnt}
            \For{each minimal generalization $CandP$ of $P$ by
              time}
            \State Lines \ref{lin:st} to \ref{lin:end} above 
            \EndFor
         \EndIf  
     \EndFor
     \State $\ell$ = $\ell$ + 1       
\EndWhile                                 
\end{algorithmic}
\caption{Optimized Algorithm for enumerating rank-based ODT patterns}
\label{algo:rankbased}
\end{algorithm}

Based on the above lemmas, we can prove the correctness of our
enumeration algorithm for rank-based ODT patterns, described by
Algorithm \ref{algo:rankbased}.
The algorithm computes first all level-$3$ patterns $\mathcal{P}_3$, based on the
atomic pattern support threshold $s_a$ (Lines 3--5).
Having the patterns at level $\ell$, the algorithm organizes those at
level $\ell + 1$ in a priority queue (minheap)  $\mathcal{P}_{\ell+1}$ of
maximum size $k$.
We consider all patterns $P$ at level $\ell$ in decreasing order of
support  $P$.cnt, to maximize the potential of generating level-$(\ell
+ 1)$ triples of high support early.
For each such pattern $P$, we first check if $P$ can
generate any level-$(\ell+1)$ triple that can enter the set $\mathcal{P}_{\ell+1}$ of
top-$k$ triples so far at level $\ell+ 1$, based on Lemma
\ref{lem:prunep}.
If this is not possible, then $P$ is pruned.
Otherwise, we attempt to generalize $P$, first by adding an atomic
region to $P.O$. If the maximum addition to $P$.cnt by such an
extension cannot result in a $CandP$ that can enter the top-$k$ at
level $\ell+ 1$ (based on Lemma \ref{lemma:maxsupport}), then we do
not attempt such extensions; otherwise we try all such extensions and
measure their supports (Lines \ref{lin:st} to \ref{lin:end}).
We repeat the same for the possible extensions of $P.D$ and $P.T$.
After $\mathcal{P}_{\ell+1}$ has been finalized, we use it to generate
the top-$k$ patterns at the next level.
Since the number of levels for which we can generate patterns can be
very large, Algorithm \ref{algo:rankbased} takes as a parameter the
maximum level $maxl$ for which we are interested in generating patterns.

\section{Approximate Computation and Sampling of ODT patterns}\label{sec:approximate2}
Due to the bottom-up, Apriori-style generation of candidates, the enumeration of large ODT patterns can be  expensive.
The combinations of atomic patterns that can formulate generalized ones exponentially increase as the generalized origins, destinations, and timeslots increase in size.
The techniques described in the previous section reduce the cost of pattern enumeration by placing constraints or by ranking the ODT triples at each level and considering only the ones of highest support as patterns; hence, limiting the number of candidates at each level. Rank-based ODT patterns are essentially an approximate solution, which, however, does not avoid the potentially expensive level-wise pattern generation while risking missing some important patterns.

In this section, we investigate an alternative approximate solution to the pattern enumeration problem, which does not apply level-wise generation of candidate ODT patterns. Our proposal is a randomized solution, which targets the detection of ODT patterns of a certain size. Specifically, given a target size $S_O$ of generalized origins $O$, a target size $S_D$ of $D$, and a target size $S_T$ of $T$, our approach efficiently generates-and-tests random ODT triples and keeps the ones that satisfy the support threshold $s_r$.
Specifically, to generate a random candidate ODT pattern, our algorithm picks a random node $v\in V$ as seed and performs BFS around $v$ to generate a connected subgraph with $S_O$ nodes. This process is repeated  $|V|$ times to generate $|V|$ subgraphs which form a set of O-candidates $\mathcal{S}_O$. We repeat the same to generate the D-candidates $\mathcal{S}_D$. Then, we pick random pairs of O- and D-candidates and complete them with random periods of $S_T$ consecutive timeslots, to generate  candidate ODT patterns. For each such candidate ODT-pattern, we count its support and add it to the set of approximate patterns if the support is greater than $s_r$.

\subsection{Description and analysis}\label{sec:approxf}
\begin{figure}
  \centering
    \includegraphics[width=0.7\textwidth]{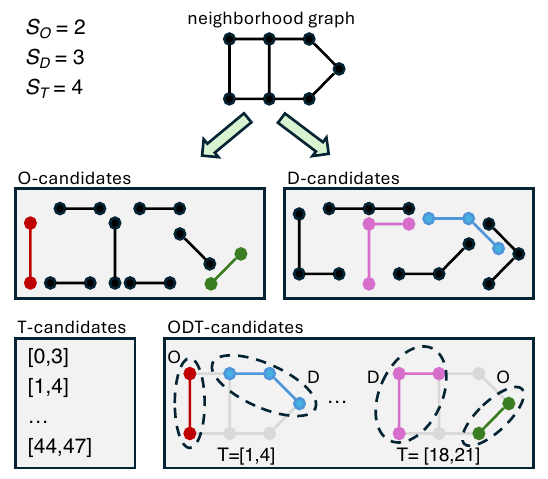}
    \caption{Illustration of the approximate algorithm. Suppose that
      we are looking for patterns with 2 regions in the origin
      ($S_O=2$), 3 regions in the destination ($S_D=3$), and 4
      timeslot ($S_T=4$). We apply BFS from each node in the region neighborhood graph to generate candidate O and D components of the desired sizes
      and randomly pick them and combine them with random T periods of 4 timeslots to generate candidate ODT triples, which are eventually verified.}
  \label{fig:approx2}
\end{figure}

Figure \ref{fig:approx2} illustrates the process and Algorithm \ref{algo:approxf} presents its steps in detail. We first initialize the two sets $\mathcal{S}_O$ and $\mathcal{S}_D$ with the O-candidates and D-candidates, respectively.
For each node $v\in V$, we place $v$ in a new set $S$ and perform a {\em random} BFS around $v$ to form a random connected subgraph centered at $v$ having exactly $S_O$ nodes. To achieve this, we initialize a FIFO queue $Q$ with the neighbors of $v$ in the region neighborhood graph, in random order. At each step of the while loop (Lines 5--10), we dequeue the next node $u$ from $Q$ and if it is not part of $S$ we add it $u$ to $S$ and its neighbors to $Q$ in random order. As soon as $S$ has $S_O$ nodes, we terminate and add $S$ to the set $\mathcal{S}_O$ of the O-candidates. The nodes in $S$ are guaranteed to form a connected subgraph, due to the BFS. Randomness in adding the neighbors guarantees that $S$ is formed by a random subtree sized $S_O$, rooted at $v$, since there could be multiple such subtrees.
The same BFS process is repeated to form a D-candidate from $v$ to be added to $\mathcal{S}_D$ (Lines 12--20).
The time required to generate $\mathcal{S}_O$ and $\mathcal{S}_D$ is $O(|V|S_O)$ and $O(|V|S_D)$, respectively. 
Next, we collect into set $\mathcal{T}$ all T-candidates as all subsequences of $S_O$ consecutive atomic timeslots, costing $O(|\mathcal{T}|)$ time, where $|\mathcal{T}|$ is the number of atomic timeslots.
Finally (Lines 27--41), the algorithm picks $M$ combinations of
(O-candidate, D-candidate,T-candidate) triples $(O,D,T)$ having $O\cap
D\ne \emptyset$ (as per the requirement of ODT patterns), and counts
their supports given the known set of atomic patterns. Recall that
atomic patterns are the top $s_a$ fraction of atomic triples by
support, computed initially  (at
Line 1). If the ratio \textit{cnt} $/$ ($|O|$ $\times$ $|D|$ $\times$ $|T|$) is at least $s_r$, then ($O, D, T$) is added to the set of discovered patterns (Lines 60-62).
The algorithm terminates after $M$ sampling iterations and returns all discovered patterns.

Each randomly generated candidate ODT pattern includes  $S_O\cdot S_D\cdot S_T$ atomic triples, hence counting the supports of $M$ candidate ODT patterns to form the final set of results (Patterns) costs $O(M\cdot S_O\cdot S_D\cdot S_T)$. 
The space requirements of the algorithm are minimal, as we only
generate one O-candidate and one D-candidate per graph node. In
addition, we only generate at most one T-candidate per atomic
timeslot. Hence, the space complexity is $O((S_O+S_D)\cdot |V|+|\mathcal{T}|)$. 


\begin{algorithm}
\scriptsize
\begin{algorithmic}[1]
\Require Graph $G(V,E)$, trips data, $s_a$, $s_r$, $S_O$, $S_D$, $S_T$, $M$
\State Compute atomic patterns (top $s_a$ fraction by support)

\Comment Step 1: Generate O and D candidates

\State $\mathcal{S}_O \gets \emptyset$, $\mathcal{S}_D \gets \emptyset$; Patterns $\gets \emptyset$
\For{each node $v \in V$}
    \State $S \gets \{v\}$;
    Initialize FIFO queue $Q$ with neighbors of $v$, pushed to $Q$ in random order
    
    \While{$Q \neq \emptyset$ and $|S| < S_O$}
        \State $u \gets \text{dequeue}(Q)$
        \If{$u \notin S$}
            \State $S \gets S \cup \{u\}$;
            Push neighbors of $u$ to $Q$ in random order
        \EndIf
    \EndWhile
    \State Add $S$ to $\mathcal{S}_O$ \Comment{new O-candidate}
    \State $S \gets \{v\}$
    \State Initialize FIFO queue $Q$ with neighbors of $v$, pushed to $Q$ in random order
    
    \While{$Q \neq \emptyset$ and $|S| < S_D$}
        \State $u \gets \text{dequeue}(Q)$
        \If{$u \notin S$}
        \State $S \gets S \cup \{u\}$;
        Push neighbors of $u$ to $Q$ in random order
        \EndIf
    \EndWhile
    \State Add $S$ to $\mathcal{S}_D$ \Comment{new D-candidate}

    
\EndFor

\Comment Step 2: Generate $T$ candidates

\State $\mathcal{S}_T \gets \emptyset$
\State Let $|\mathcal{T}|$ be the total number of timeslots
\For{each integer $i \in [0, |\mathcal{T}| - S_T]$}
    \State Add timeslot interval $[i, i + S_T - 1]$ to $\mathcal{S}_T$
\EndFor

\Comment Step 3: Random sampling for ODT patterns

\For{$i = 1$ to $M$}\Comment{Sample $M$ random combinations}
    \State Pick random $O \in \mathcal{S}_O$, $D \in \mathcal{S}_D$, $T \in \mathcal{S}_T$
    \If{$O \cap D \neq \emptyset$}
    \State $i \gets i-1$;
    {\bf continue} \Comment{Repeat picking} 
    \EndIf
    
    \State $\text{cnt} \gets 0$
    \For{each $(o,d,t) \in O \times D \times T$}
        \If{$(o,d,t)$ is an atomic pattern}
            \State $\text{cnt} \gets \text{cnt} + 1$
        \EndIf
    \EndFor
    
    \If{$\text{cnt} / (|O| \times |D| \times |T|) \geq s_r$}
        \State Add $(O, D, T)$ to Patterns
    \EndIf
\EndFor
\State \Return Patterns
\end{algorithmic}
\caption{Approximate Algorithm for ODT Pattern Discovery}\label{algo:approxf}
\end{algorithm}

\subsection{A weighted approximate ODT pattern detection variant}\label{sec:approxfw}
Our approximate algorithm samples ODT candidates with equal probability.
An improved version would pick ODT candidates that have higher probability to be ODT patterns. For this, the O- D-, and T-candidates in $\mathcal{S}_O$,  $\mathcal{S}_D$ and $\mathcal{S}_T$, respectively, are {\em weighted} and the $M$ ODT pattern candidates are selected considering these weights. For the weighting, we use the following definitions:

\begin{enumerate}
\item For each node $v\in V$, we define as weight $w_O(v)$ the total number of atomic ODT patterns which include $v$ as origin; we define as $w_D(v)$ the total number of atomic ODT patterns which include $v$ as destination.
\item For each atomic timeslot $t\in \mathcal{T}$, we define as weight $w_T(t)$ the total number of atomic ODT patterns which include $t$.
\item For each set of nodes $S\subseteq V$, we define  as weight $w_O(S)$ the sum $\sum_{v\in S}w_O(v)$ and as $w_D(S)$ the sum $\sum_{v\in S}w_D(v)$.
  Hence, each set $S$ of nodes, which is an O-candidate in $\mathcal{S}_O$ or a D-candidate in $\mathcal{S}_D$ has a weight. 
\item For each sequence $T$ of consecutive atomic timestamps in $\mathcal{T}$ which is a T-candidate, we define  as weight $w_T(T)$ the sum $\sum_{t\in T}w_T(t)$.
\end{enumerate}

Intuitively, combinations of O-, D-, and T-candidates with larger combined weight have higher probability to be an ODT pattern. Hence, we apply weighted ranking of the ODT candidates and use the top $M$ candidates instead of uniformly sampling at random from the pool of ODT candidates.
Algorithm \ref{algo:approxfw} describes this weighted approximate ODT pattern detection method.
First, we compute the $w_O(v)$ and $w_D(v)$ weights for each node $v$ in the region neighborhood graph $V$. Then, we run Lines 3--21 of Algorithm \ref{algo:approxf} to compute the O-candidates and D-candidates as well as their respective $w_O$ and $w_D$ weights. The next step is to compute the weights of atomic timeslots and the T-candidates and their $w_T$ weights.
Finally, we enumerate all ODT candidates (combinations) and for each
one we set its combined weight as the product of the weights of its
O-, D-, and T- components (Line 13). While doing so, we update in a
min-heap the top-$M$ ODT candidates based on combined weights. In
Lines 17--28, the supports of all these ODT candidates are computed
and the ones which are ODT patterns are added to the Patterns result
set.

\stitle{Complexity Analysis}
Algorithm \ref{algo:approxfw} initially scans once the set $\mathcal{P}_3$ of atomic
patterns to generate the weights of the atomic regions (i.e., graph
nodes)
and the weights of
the atomic timeslots.
This costs $O(|\mathcal{P}_3|)$ time.
Then, the O-candidates, D-candidates, and T-candidates are determined
in $O(|V|S_O+|V|S_D+|\mathcal{T}|)$ time, as already discussed for 
Algorithm \ref{algo:approxf}.
Then, the algorithm enumerates of all ODT
candidates for the determination of the top-$M$ candidates to be
verified.
This costs $O(S_O\cdot S_D\cdot S_T\cdot \log M)$ with the help of the
priority queue (heap) $H$.
Finally, the verification of the candidates costs
$O(|\mathcal{P}_3|\cdot M)$.
Overall, the total cost is $O(S_O\cdot S_D\cdot S_T\cdot \log M +
|\mathcal{P}_3|\cdot M)$; if $M$ is large compared to the number
of ODT candidates, the
$|\mathcal{P}_3|\cdot M)$ factor dominates. 
The space complexity is just $O(|\mathcal{P}_3| +M)$ as we only need to
store the atomic ODT patterns and the set of top-$M$ ODT candidates to
be verified.

\begin{algorithm}
\scriptsize
\begin{algorithmic}[1]
\Require Graph $G(V,E)$, trips data, $s_a$, $s_r$, $S_O$, $S_D$, $S_T$, $M$
\State Compute atomic patterns (top $s_a$ fraction by support)
\State For each node $v\in V$, use the atomic patterns to compute $w_O(v)$ and $w_D(v)$

\Comment Step 1: Generate O and D candidates
\State $\mathcal{S}_O \gets \emptyset$, $\mathcal{S}_D \gets \emptyset$; Patterns $\gets \emptyset$
\State Same as Lines 3--21 of Algorithm \ref{algo:approxf}
\State For each O-candidate $S\in \mathcal{S_O}$ compute $w_O(S)$ 
\State For each D-candidate $S\in \mathcal{S_D}$ compute $w_D(S)$ 

\State For each node $t\in \mathcal{T}$, use the atomic patterns to define $w_T(t)$

\Comment  Step 2: Generate $T$ candidates
\State Same as Lines 22--26 of Algorithm \ref{algo:approxf}

\State For each T-candidate $S\in \mathcal{S_T}$ compute $w_T(T)$

\Comment Step 3: Examine ODT candidates in decreasing order of combined weight

\State $H \gets$ empty min-heap, to hold top-$M$ ODT candidates by combined weight
\For{all $ODT$, $O \in \mathcal{S}_O$, $D \in \mathcal{S}_D$, $T \in \mathcal{S}_T$}
  \If{$O \cap D = \emptyset$}
  \State $w(ODT)\gets w_O(O)\cdot w_D(D)\cdot w_T(T)$
  \State Update $H$ to hold top-$M$ ODTs by $w(ODT)$
  \EndIf
\EndFor

\For{$i = 1$ to $M$}\Comment{For each ODT candidate in $H$}
    \State $ODT\gets H[i]$
    \State $\text{cnt} \gets 0$
    \For{each $(o,d,t) \in O \times D \times T$}
        \If{$(o,d,t)$ is an atomic pattern}
            \State $\text{cnt} \gets \text{cnt} + 1$
        \EndIf
    \EndFor
    
    \If{$\text{cnt} / (|O| \times |D| \times |T|) \geq s_r$}
        \State Add $(O, D, T)$ to Patterns
    \EndIf
\EndFor
\State \Return Patterns
\end{algorithmic}
\caption{Weighed Approximate Algorithm for ODT Pattern Discovery}\label{algo:approxfw}
\end{algorithm}

\section{Experimental Evaluation}\label{sec:exps}
In this section, we evaluate the performance of our proposed
algorithms on real datasets.
All methods were implemented in Python 3.14 and the experiments were run on
a Macbook Pro with a M4 processor and 16GB memory. The source code of
the paper is publicly
available\footnote{https://github.com/kosyfakicse/Multi-granularity-ODT-patterns-Enumeration}.

\subsection{Dataset Description}
For our experiments, we used three real datasets; NYC
taxi trips, MTR network trips
and Flights.
Below,
we provide a detailed description for each of them.

\stitle{NYC taxi trips:} We processed 7.5M trips of yellow taxis in NYC
in January 2019, downloaded from TLC%
\footnote{https://www.nyc.gov/site/tlc/about/tlc-trip-record-data.page}.
Each record represents a taxi trip and
includes the pick-up and drop-off taxi zones
(different regions in NYC), the date/time of the pick-up, and the
number of passengers who took the trip.
The 315 taxi zones represent atomic regions. 
We converted all time moments to 48 time-of-day slots (one slot per
30min intervals in the 24h). Then, we aggregated the data by merging
all trips having  the same origin, destination, and
timeslot, and summing up the total number of passengers in all these
trips to a total passenger flow, as explained in Section
\ref{sec:def}. This way, we ended up having 373460 unique ODT combinations
 (atomic ODT triples), which we used as input to our pattern
 enumeration algorithms.
In addition, we used the maps posted at the same website to construct
the neighboring graph $G$ between the atomic regions (taxi zones).
In $G$, we connected all pairs of atomic regions that share boundary
points or are separated by water boundaries.

\stitle{MTR trips:} The Mass Transit Railway (MTR)\footnote{https://www.mtr.com.hk/en/customer/main/index.html} is the biggest
and the most popular public transport network operating in Hong
Kong. The system consists of 168 stations, serving the areas of Hong
Kong island, Kowloon and New Territories.
We consider each station as an atomic region; we created the
neighborhood graph $G$ for them by linking stations that are next to
each other in the network.
MTR Corp. provided us with aggregate data for all passenger trips
taken in September 2019. Specifically, for each atomic ODT triple,
where the origin and destination are MTR stations and T is one of the 
48 atomic timeslots, we have the total number of passenger trips
in Sep. 2019.
The total number of  atomic ODT triples is 253497.

\stitle{Flights:} We extracted information for 5.8M US flights in 2015 from Kaggle%
\footnote{https://www.kaggle.com/datasets/usdot/flight-delays?select=flights.csv}.
In
this dataset, we consider as atomic regions 319 airports in North
America that appear in the file.
Since the number of passengers in each flight was not given in the original data, we randomly generated a number between 50 and 200.
We followed the same procedure as in for the two previous
datasets; namely, we converted the original flights data into a table with
atomic ODT triples.
The total number of resulting ODT triples is 17623.
To
create the neighbor graph $G$, we
follow the same logic as the two previous datasets; we connect
atomic regions in neighboring states.

\begin{figure*}[t!]
  \centering
 \subfigure[Taxi Network]{
   \label{fig:exp:sataxi}
    \includegraphics[width=0.30\textwidth]{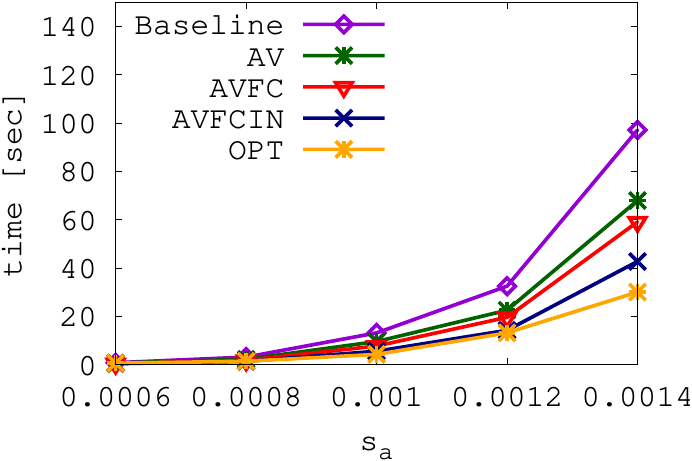}
    }\!\!
  \subfigure[MTR Network]{
   \label{fig:exp:samtr}
    \includegraphics[width=0.30\textwidth]{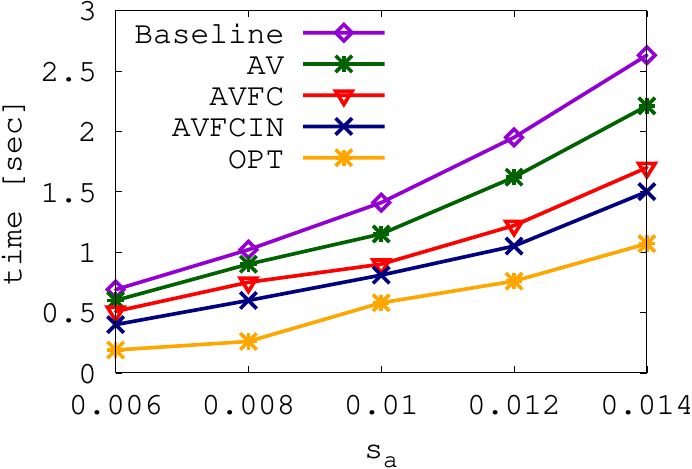}
    }\!\!
  \subfigure[Flights Network]{
   \label{fig:exp:saflights}
   \includegraphics[width=0.30\textwidth]{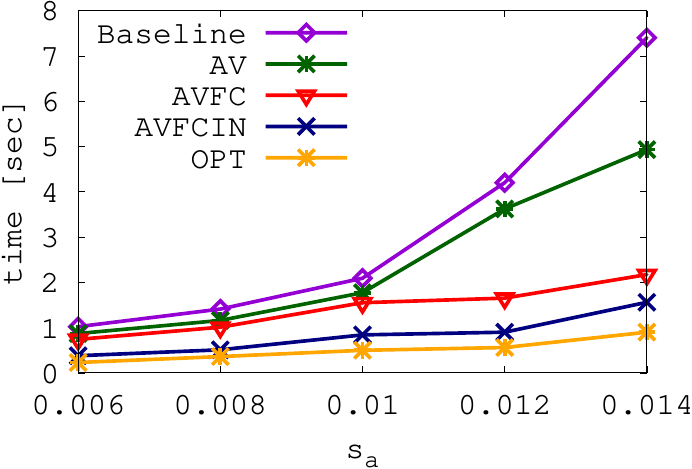}
    }
  \caption{Pattern enumeration runtime, $s_r=0.5$, varying $s_a$}
  \label{fig:exp:time}
\end{figure*}

\begin{figure*}[t!]
  \centering
 \subfigure[Taxi Network]{
   \label{fig:exp:taxitime}
    \includegraphics[width=0.30\textwidth]{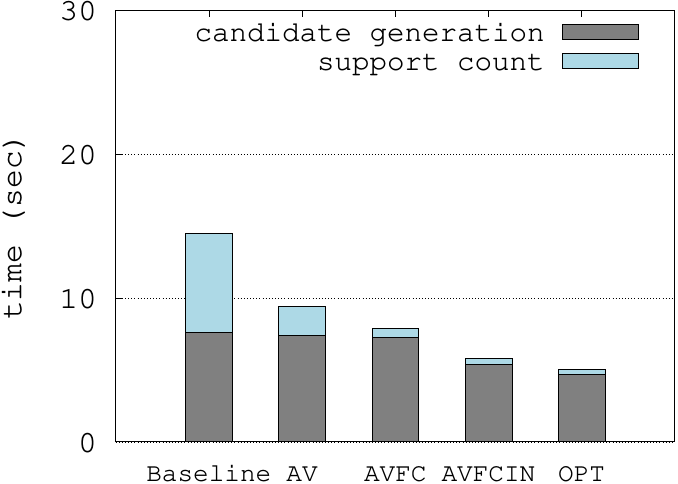}
    }\!\!
  \subfigure[MTR Network]{
   \label{fig:exp:mtrtime}
    \includegraphics[width=0.30\textwidth]{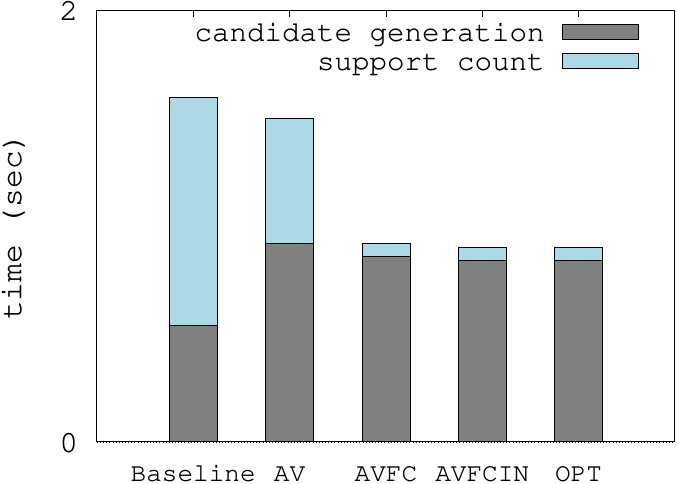}
    }\!\!
  \subfigure[Flights Network]{
   \label{fig:exp:flightstime}
   \includegraphics[width=0.30\textwidth]{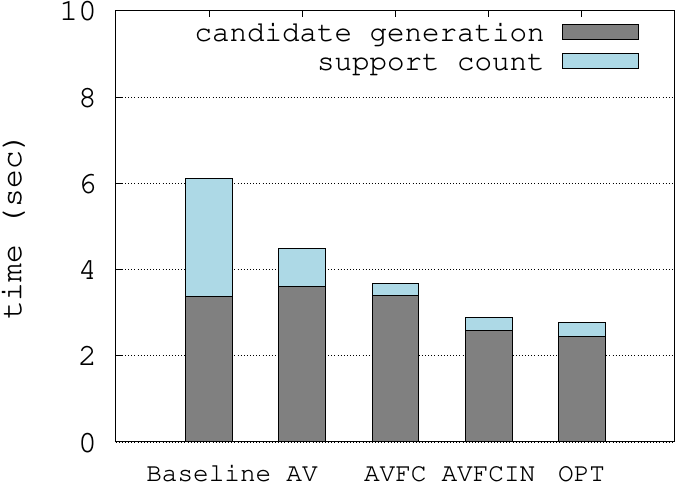}
    }
  \caption{Pattern enumeration cost breakdown, $s_r=0.5$, default $s_a$}
  \label{fig:exp:timebreakdown}
\end{figure*}

\begin{figure*}[t!]
  \centering
 \subfigure[Taxi Network]{
   \label{fig:exp:srtaxi}
    \includegraphics[width=0.30\textwidth]{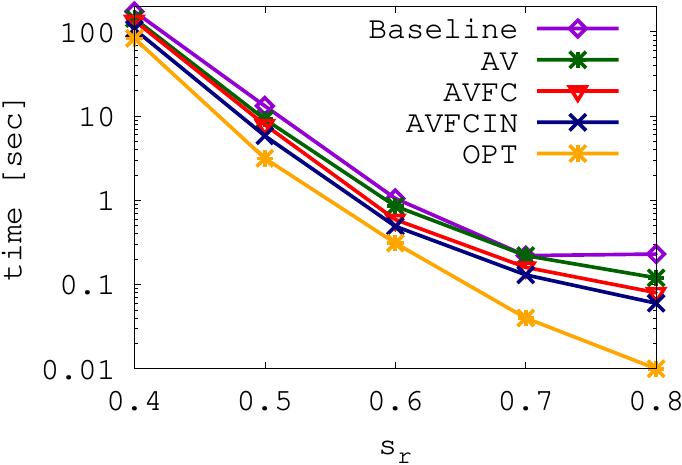}
    }\!\!
  \subfigure[MTR Network]{
   \label{fig:exp:srmtr}
    \includegraphics[width=0.30\textwidth]{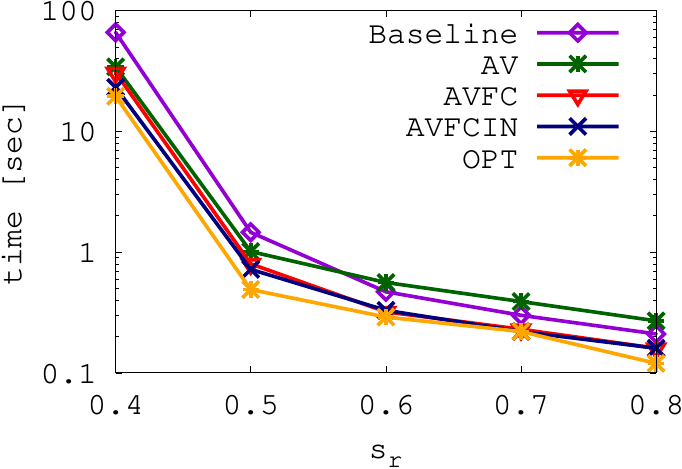}
    }\!\!
  \subfigure[Flights Network]{
   \label{fig:exp:srflights}
   \includegraphics[width=0.30\textwidth]{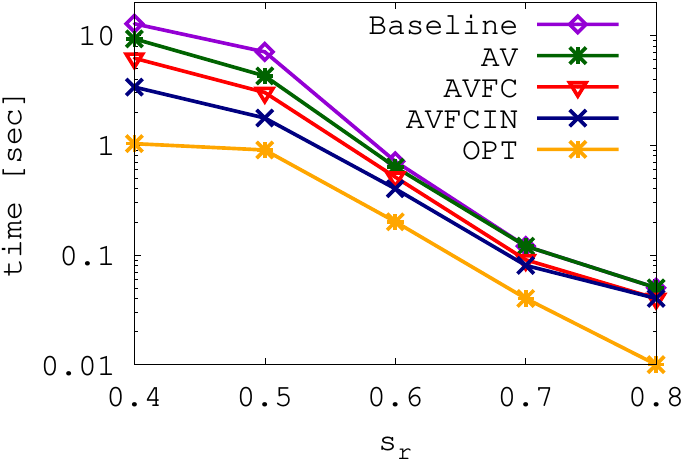}
    }
  \caption{Pattern enumeration runtime, default $s_a$, varying $s_r$}
  \label{fig:exp:sr}
\end{figure*}

\begin{figure*}[t!]
  \centering
 \subfigure[Taxi Network]{
   \label{fig:exp:taxipatterns}
    \includegraphics[width=0.30\textwidth]{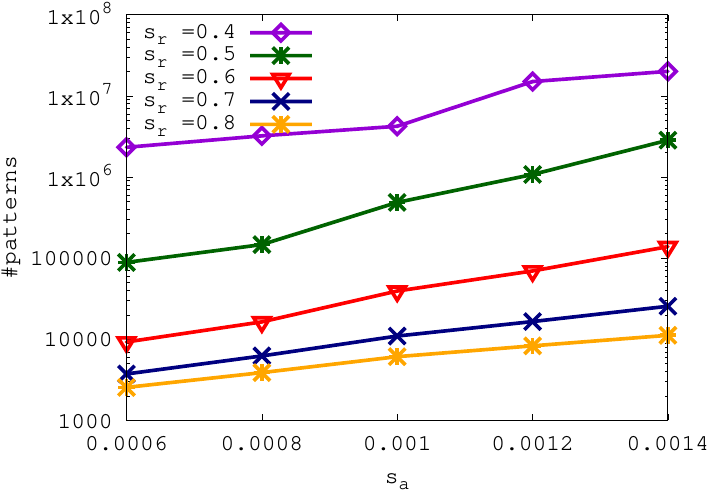}
    }\!\!
  \subfigure[MTR Network]{
   \label{fig:exp:mtrpatterns}
    \includegraphics[width=0.30\textwidth]{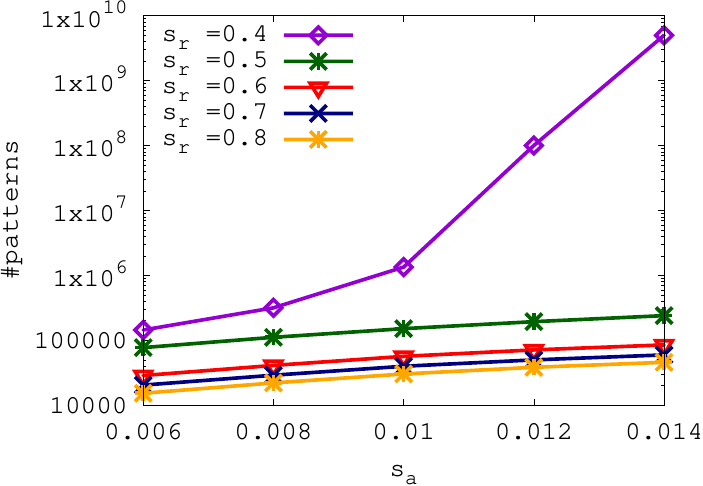}
    }\!\!
  \subfigure[Flights Network]{
   \label{fig:exp:flightspatterns}
   \includegraphics[width=0.30\textwidth]{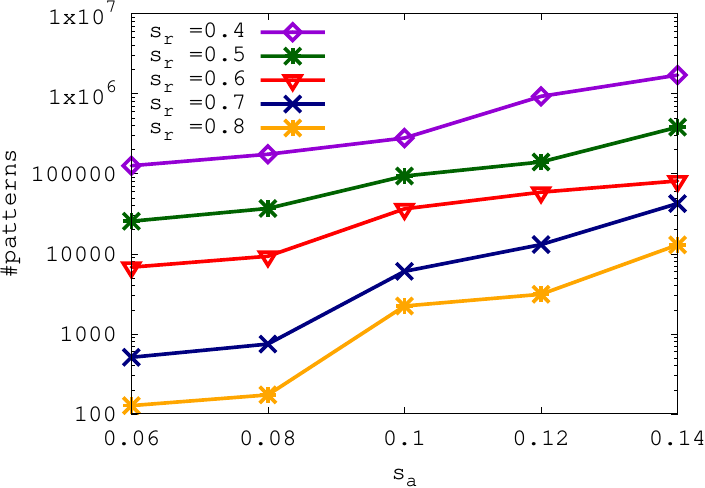}
    }
  \caption{Number of patterns for different values of $s_a$ and $s_r$}
  \label{fix:exp:patterns}
\end{figure*}


\begin{figure*}[t!]
  \centering
 \subfigure[Taxi Network]{
   \label{fig:exp:orgtaxi}
    \includegraphics[width=0.30\textwidth]{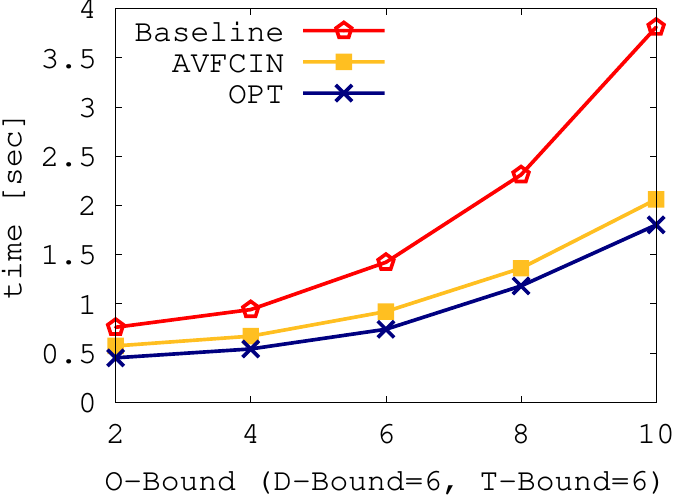}
    }\!\!
  \subfigure[MTR Network]{
   \label{fig:exp:orgmtr}
    \includegraphics[width=0.30\textwidth]{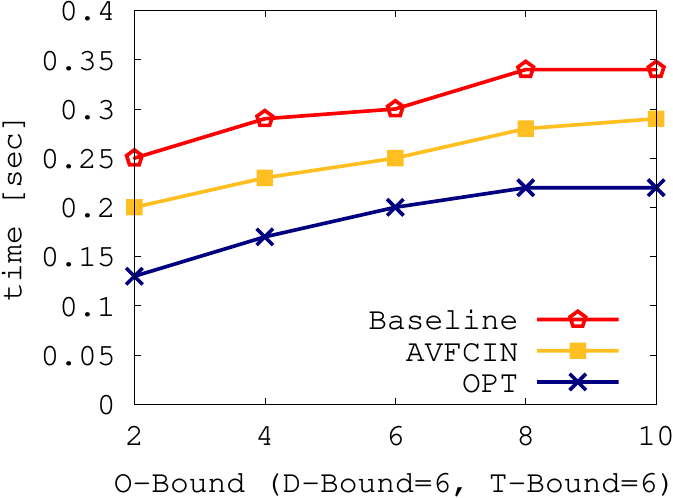}
    }\!\!
  \subfigure[Flights Network]{
   \label{fig:exp:orgflights}
   \includegraphics[width=0.30\textwidth]{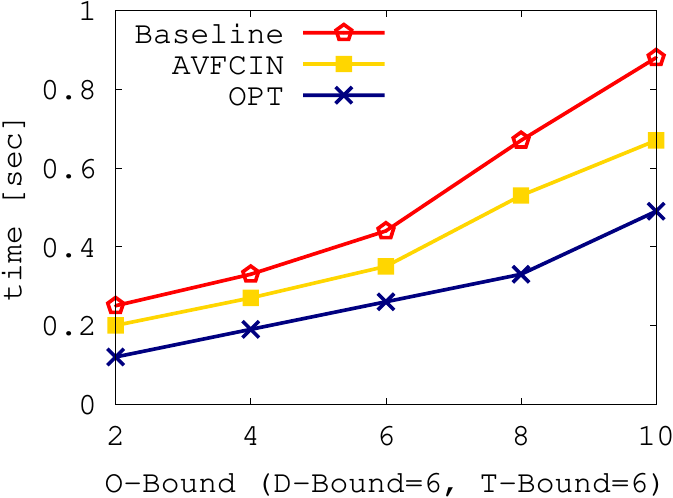}
    }
  \caption{Bounded pattern enumeration runtime, default $s_a$, $s_r$,
    varying origin bound}
  \label{fig:exp:src}
\end{figure*}

\begin{figure*}[t!]
  \centering
 \subfigure[Taxi Network]{
   \label{fig:exp:desttaxi}
    \includegraphics[width=0.30\textwidth]{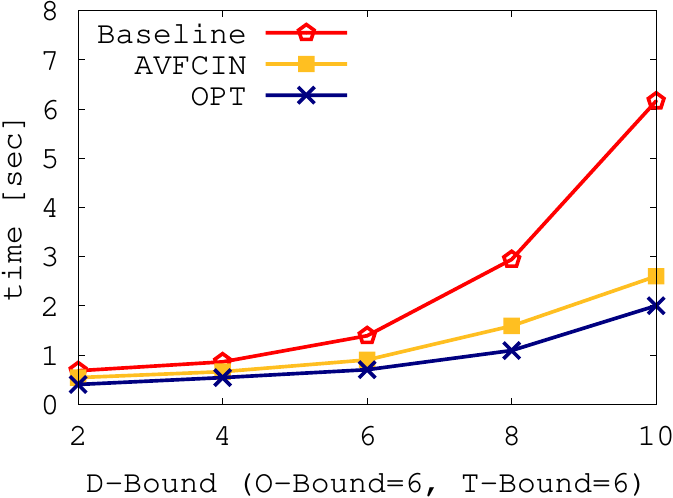}
    }\!\!
  \subfigure[MTR Network]{
   \label{fig:exp:destmtr}
    \includegraphics[width=0.30\textwidth]{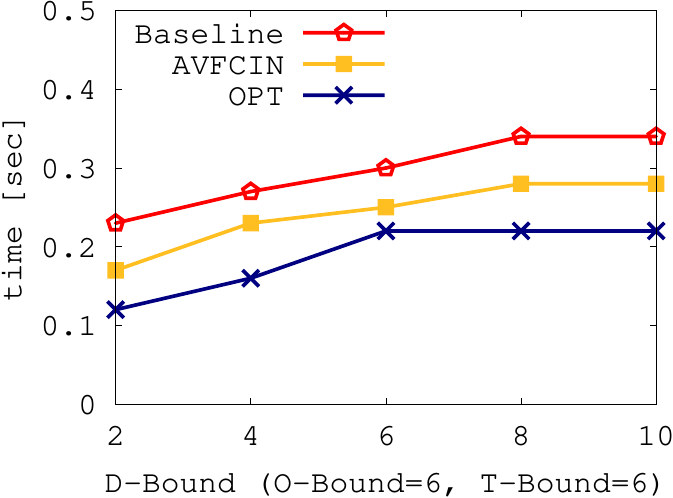}
    }\!\!
  \subfigure[Flights Network]{
   \label{fig:exp:destflights}
   \includegraphics[width=0.30\textwidth]{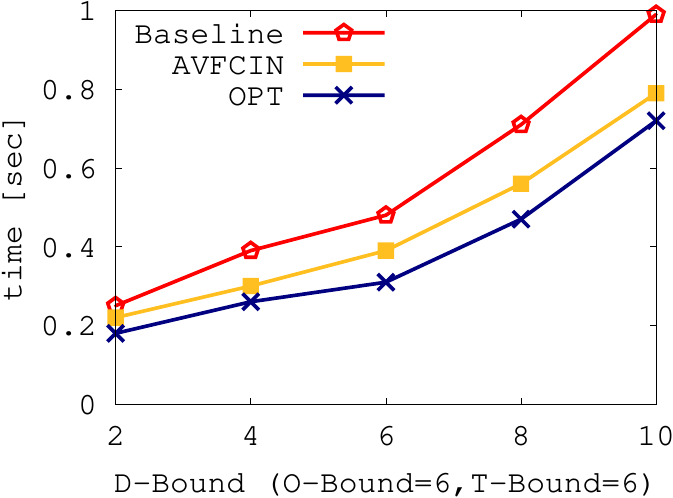}
    }
  \caption{Bounded pattern enumeration runtime, default $s_a$, $s_r$,
    varying destination bound}
  \label{fig:exp:dest}
\end{figure*}

\begin{figure*}[t!]
  \centering
 \subfigure[Taxi Network]{
   \label{fig:exp:timetaxi}
    \includegraphics[width=0.30\textwidth]{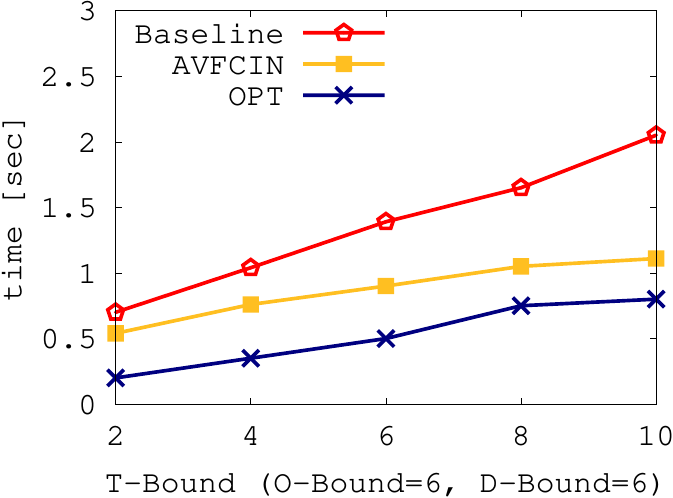}
    }\!\!
  \subfigure[MTR Network]{
   \label{fig:exp:timemtr}
    \includegraphics[width=0.30\textwidth]{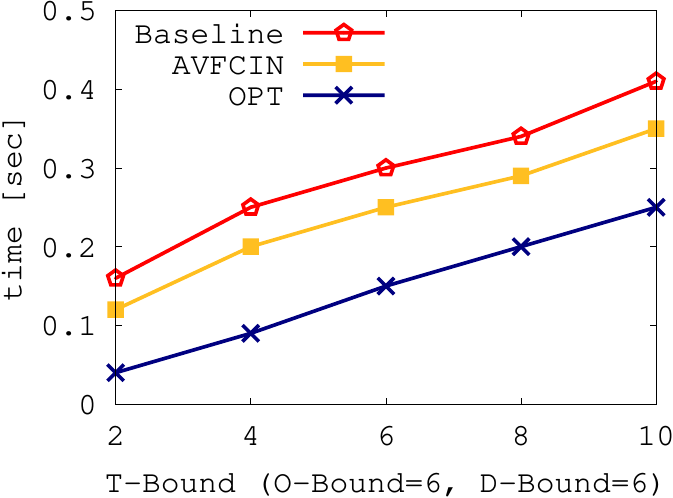}
    }\!\!
  \subfigure[Flights Network]{
   \label{fig:exp:timeflights}
   \includegraphics[width=0.30\textwidth]{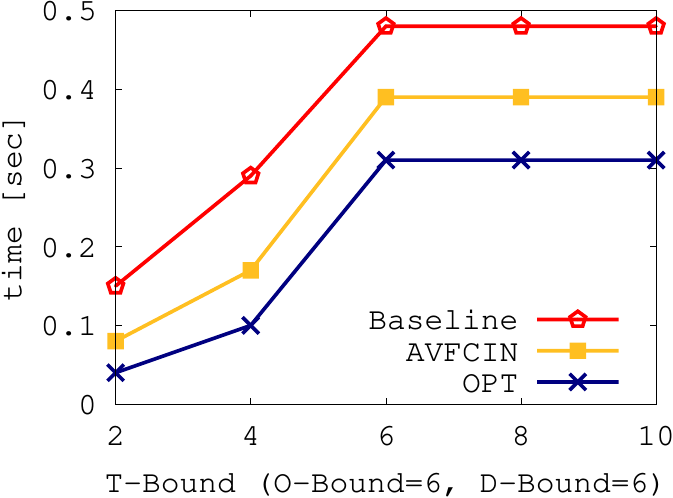}
    }
  \caption{Bounded pattern enumeration runtime, default $s_a$, $s_r$,
    varying timeslot bound}
  \label{fig:exp:timec}
\end{figure*}


\begin{figure*}[t!]
  \centering
 \subfigure[Taxi Network]{
   \label{fig:exp:taxilevel}
    \includegraphics[width=0.30\textwidth]{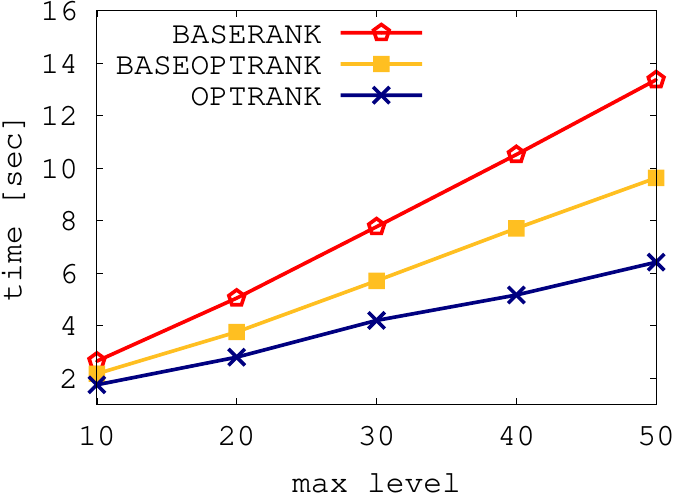}
    }\!\!
  \subfigure[MTR Network]{
   \label{fig:exp:mtrlevel}
    \includegraphics[width=0.30\textwidth]{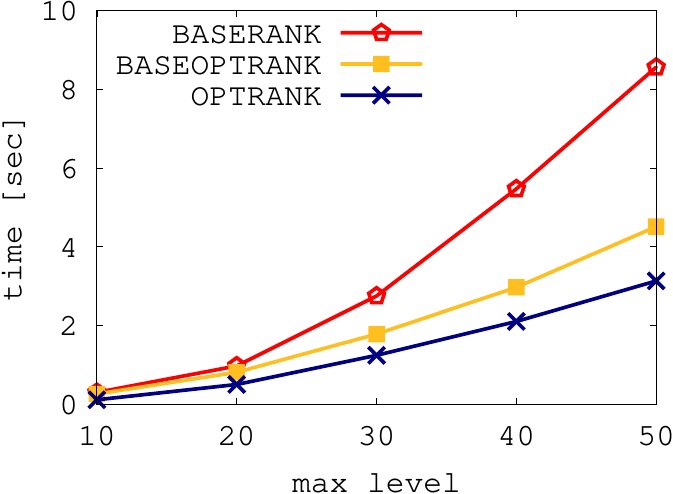}
    }\!\!
  \subfigure[Flights Network]{
   \label{fig:exp:flightslevel}
   \includegraphics[width=0.30\textwidth]{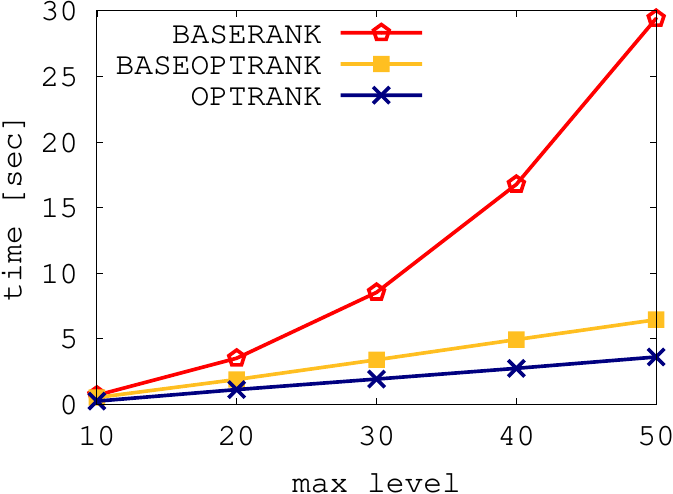}
    }
  \caption{Rank-based pattern enumeration, $s_a=0.1$, $k=3000$, varying $maxl$}
  \label{fig:exp:level}
\end{figure*}

\begin{figure*}[t!]
  \centering
 \subfigure[Taxi Network]{
   \label{fig:exp:taxik}
    \includegraphics[width=0.30\textwidth]{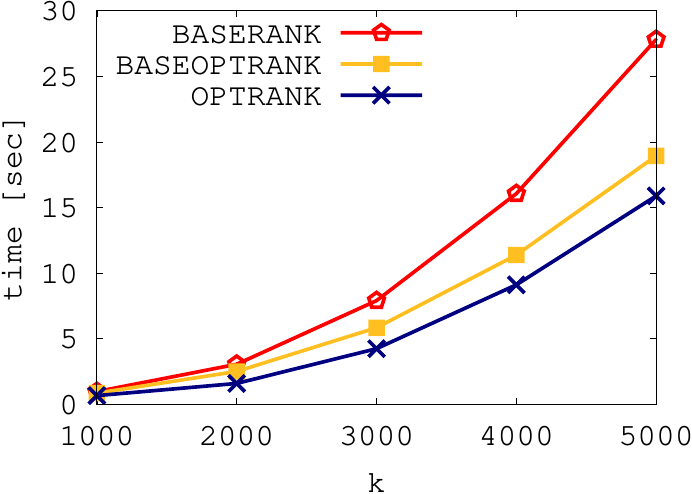}
    }\!\!
  \subfigure[MTR Network]{
   \label{fig:exp:mtrk}
    \includegraphics[width=0.30\textwidth]{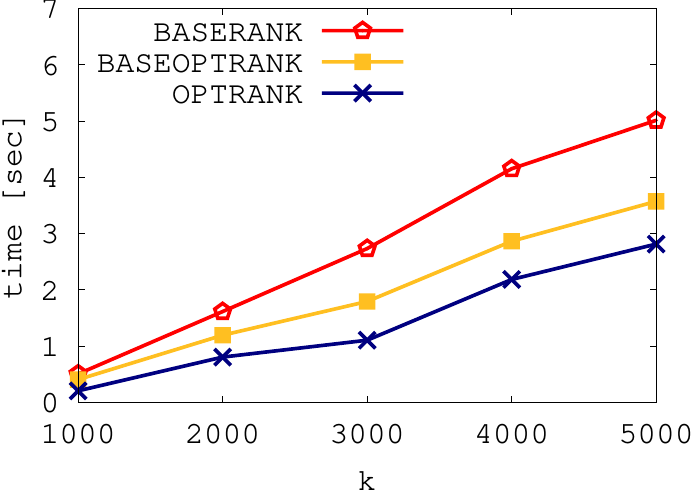}
    }\!\!
  \subfigure[Flights Network]{
   \label{fig:exp:flightsk}
   \includegraphics[width=0.30\textwidth]{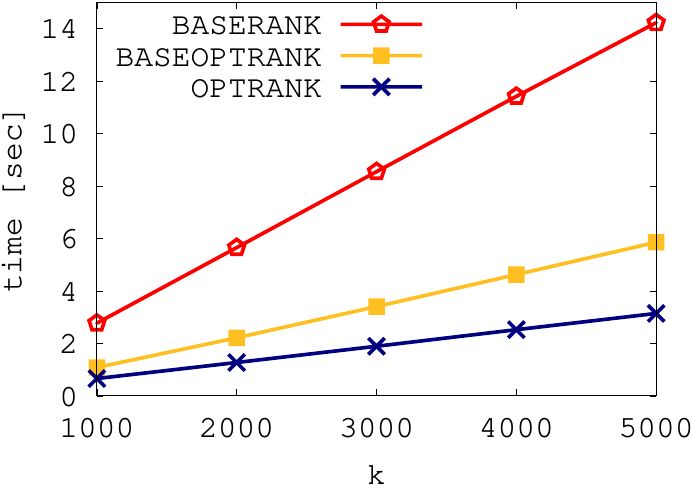}
    }
  \caption{Rank-based pattern enumeration, $s_a=0.1$, $maxl=30$, varying $k$}
  \label{fig:exp:k}
\end{figure*}

\subsection{Pattern enumeration}
We start by evaluating the performance of our
baseline pattern enumeration algorithm, described in Section
\ref{sec:baseline}, and its optimizations, described in Section \ref{sec:opt}.
Specifically, we compare the performance of the following methods:

  \begin{itemize}
  \item Algorithm \ref{algo:baseline}, denoted by Baseline.
  \item Algorithm \ref{algo:baseline}  with the avoid recounting $P'$
    optimization, denoted by AV. 
    \item Algorithm \ref{algo:baseline}  with the avoid recounting $P'$
      and fast check for zero support of $P'$ optimizations, denoted by AVFC.
    \item Algorithm \ref{algo:baseline}  with the avoid recounting $P'$,
      fast check, and improved neighborhood optimizations, denoted by AVFCIN.
     \item Algorithm \ref{algo:baseline}  with all four optimizations,
       denoted by OPT.
    \end{itemize}

Figure \ref{fig:exp:time} shows the costs of all tested methods on the three datasets
for various values of $s_a$ (default $s_a=0.001$ for Taxi, 
$s_a=0.01$ for MTR and Flights), while keeping $s_r$ fixed to 0.5. Observe
that the optimizations pay off, since the initial cost of the baseline
approach drops to about 50\% of the initial cost. When comparing
between the different optimizations, we observe that the ones that
have the biggest impact are the $P'$ counting avoidance and the
improved neighborhood computation. The savings by the prefix sum
optimization are not impressive, because the
other optimizations already reduce a lot the number of candidates for which
exact counting is required.

This assertion is confirmed by the cost-breakdown experiment shown in
Figure \ref{fig:exp:timebreakdown}, where for the default values of $s_a$
and $s_r$, we show the fraction of the cost that goes to candidate
pattern generation and support counting. Note that the baseline
approach spends most of the time in pattern counting, as the candidate
generation process is quite simple. On the other hand, the optimized
versions of the algorithm trade off time for pattern generation (spent
on bookkeeping all generated triples at each level, bookkeeping OD pairs
with at least one trip, etc.) to reduce the time spent on support
counting. Note that the ratio of the time spent on support counting is
eventually minimized. When comparing between the different versions,
we observe that the candidate generation time drops as more
optimizations are employed (e.g., fast check for zero support).
Since finding all patterns at level $\ell$ requires
considering all possible extensions of patterns at level $\ell-1$, we
note that there is little room for further reducing the cost of ODT pattern
enumeration; in this respect OPT is the best approach that one can
apply if the goal is to find all ODT patterns.

Figure \ref{fig:exp:sr} shows the runtime cost of pattern enumeration
for different values of $s_r$, by keeping $s_a$ to its default
value. Observe that the cost explodes for values of $s_r$ smaller than
$0.5$. The reason is that small $s_r$ values make it easy for 
triples at each level to be characterized as patterns, which, in turn,
greatly increases the number of candidates and patterns at the next
level. On the other hand, for $s_r\ge 0.5$ at least half of the atomic
triples in a candidate must be atomic patterns, which restricts
the number of candidates and patterns at all levels.

The next experiment proves the pattern explosion for small values of
$s_r$.
The high cost of pattern enumeration stems from the fact that a very
large number of patterns are found at each level, which, in turn, all
have to be minimally generalized due to the weak monotonicity property
of Definition \ref{def:pattern}. Figure \ref{fix:exp:patterns}
shows
the numbers of enumerated patterns for different values of $s_a$ and
$s_r$. As the number of patterns grow, so does the essential cost of
candidate generation, which becomes the dominant cost factor.
From Figure \ref{fix:exp:patterns}, we observe that the number of
enumerated patterns is very sensitive to $s_r$. Specifically, for
values of $s_r$ smaller than $0.5$ the number of patterns explode. On
the other hand, the sensitivity to $s_a$ is relatively low. Still,
even for the default values of $s_a$ (0.001 for Taxi and 0.01 for MTR
and Flights) there are thousands or even millions of qualifying
patterns. Such huge numbers necessitate the use of constraints or
ranking in order to limit the number of patterns, focusing on the most
important ones.  

\stitle{Memory Requirements}
Table \ref{table:memorytaxi} shows the runtime performance (sec), the
number of patterns, and the memory requirements (GB) of all methods
for different values of $s_a$ on the Taxi dataset.
Observe that the baseline
algorithm requires significantly less memory compared to the optimized
methods, as it does not bookkeep any information (counted triples at
each level, $r.dests$ and $r.srcs$ for each atomic region $r$,
prefix-sums, etc.)
Still, the additional memory requirements by the optimized methods
are bearable and come with significant cost savings, as we have
discussed already.
In addition, large $s_a$ values that have increased memory requirements
result in a huge number of patterns, which may not be informative.

\begin{table}[ht]
  \vspace{-0.1in}
  \caption{Peak memory used varying $s_a$, default $s_r$ - Taxi Network}
  \centering
 \scriptsize
\begin{tabular}{|@{~}c@{~}|@{~}c@{~}|@{~}c@{~}|@{~}c@{~}|@{~}c@{~}|@{~}c@{~}|@{~}c@{~}| } 
\hline
$s_a$ & Performance & Baseline & AV &AVFC&AVFCIN&OPT\\
\hline
\multirow{3}{4em}{$0.001$} & \#patterns &485K&485K&485K&485K&485K\\ 
      & Memory (GB) &0.38&0.83&0.95&0.97&0.99\\
  & Time (sec) &13.1&9.5&7.6&5.5&4.9\\
\hline
\multirow{3}{4em}{$0.0012$} & \#patterns &1.07M&1.07M&1.07M&1.07M&1.07M\\ 
      & Memory (GB) &0.79&1.42&1.61&1.92&2.1\\
   & Time (sec) &32.46&22.52&19.5&14&13.2 \\
\hline
\multirow{3}{4em}{$0.0014$} & \#patterns &2.8M&2.8M&2.8M&2.8M&2.8M\\ 
      & Memory (GB) &1.1&2.72&2.88&2.99&3.4\\
   & Time (sec) &97.7&67.9&59.8&42.7&30.3 \\
  \hline
  \multirow{3}{4em}{$0.0016$} & \#patterns &23.6M&23.6M&23.6M&23.6M&23.6M\\ 
      & Memory (GB) &5.6&7.02&7.28&7.69&8.04\\
   & Time (sec) &529.4&288.9&261.1&242.6&229.8\\
\hline
\end{tabular}
\label{table:memorytaxi}
\vspace{-0.06in}
\end{table}

\subsection{Bounded patterns}

As discussed in Section \ref{sec:sizebounded}, one way to limit the
number of patterns is to bound the number of atomic regions and/or
atomic timeslots in them. In the next experiment, we study the effect
of such pattern size constraints to the runtime of algorithms
Baseline, AVFCIN, and OPT.
We run experiments by setting $s_a$ and $s_r$ to their default values.
In each experiment, we set a fixed upper bound to the sizes of two of O, D,
and T, and vary the bound of one. Hence, in Figure \ref{fig:exp:src},
we keep the upper size bounds of D and T fixed and we vary the upper size bound of
O; in Figure \ref{fig:exp:dest},
we keep the upper size bounds of O and T fixed and we vary the upper size bound of
D; in Figure \ref{fig:exp:timec},
we keep the upper size bounds of O and D fixed and we vary the upper
size bound of T. In general, the cost increases as
one bound increases, which is as expected, because the number of
patterns and generated candidates increases as well. On certain
datasets (e.g., MTR), the cost growth is slow when the bound of O or D
is increased; this is due to the fact that the number of patterns at
low levels is already quite small and the generated patterns start to
decrease as we change levels, so the bound increase does not affect
the cost significantly. On the other
hand, when the bound of T increases (Figure \ref{fig:exp:timec}),
there is a stable increase of time in all datasets. This is due to the
fact that the number of atomic timeslots is significantly small and
neighboring timeslots are highly correlated in terms of flow.
When comparing the costs of Baseline, AVFCIN, and OPT, we observe that
OPT maintains a significant performance advantage for different bound
values, especially on MTR.

\subsection{Rank-based patterns}

We now evaluate the performance of rank-based
pattern enumeration, described in Section \ref{sec:rank}.
We compare three algorithms.
The first one is the baseline approach
described in Section \ref{sec:baselinerank}, without the pattern
enumeration optimizations described in Section \ref{sec:opt}.
The second one is the baseline approach of Section
\ref{sec:baselinerank}
with the pattern
enumeration optimizations described in Section \ref{sec:opt}.
The third approach is the optimized algorithm for rank-based patterns
described in Section \ref{sec:optrank}. The three approaches are
denoted by BASERANK, BASEOPTRANK, and OPTRANK, respectively.

Figure \ref{fig:exp:level}
 shows the runtime cost of the three algorithms for
$s_a=0.1$ and $k=3000$ patterns per level, as a function of
the maximum level $maxl$ of patterns that we generate and enumerate. Recall
that the top-$k$ patterns selected per level may generate numerous
triples at the next level and there is no $s_r$ threshold to reduce
them, so the number of levels can become too large. We use $maxl$ as a
parameter for limiting the sizes of patterns.
As shown in the figure, OPTRANK maintains a large advantage over the
other approaches which do not take advantage of the pruning conditions
and the ranking of generated triples.
Figure \ref{fig:exp:k} shows the runtime cost of the algorithms for $s_a=0.1$
and various
values of $k$, after setting $maxl=30$. The advantage of  OPTRANK over
the other algorithms is not affected by $k$.
Overall, despite the fact that a very high value of $s_a$ is used, due
to the fact that the number of patterns per level is limited by $k$,
all algorithms are scalable, making pattern enumeration practical,
even in cases where the number of possible ODT combinations is huge.

\subsection{Approximate patterns}
In this section, we evaluate the efficiency and effectiveness of our
approximate algorithm
for ODT patterns detection and its weighted variant, presented in
Section \ref{sec:approximate2}. Once again,
we used our three real datasets which present different
characteristics with respect to the graph size and complexity and
with respect to the atomic ODT patterns that they may have. For
different datasets and problem parameter values,
we compare three approaches: the most efficient version of the exact
algorithm (AVFCIN
optimization), the approximate algorithm (described in Section \ref{sec:approxf}),
and its weighted variant (described in Section \ref{sec:approxfw}).
The approximate algorithms draw and verify $M=1$mil ODT candidates (either by sampling or after ranking).

\subsubsection{Effect of Varying $s_a$}

We first test the effect of the parameter $s_a$. For various values
$s_a$, Table \ref{tab:varying_sa} shows the runtime and
the number of detected patterns by the tested methods. In the
experiment, we fixed the value of $s_r=0.5$ and we are
looking for ODT patterns, where O includes 4 regions, D includes 4
regions, and T includes 4 atomic timeslots, i.e., $S_O=S_D=S_T=4$.
In some experimental instances the exact algorithm took longer than
180 seconds and it was terminated it before it could enumerate
any patterns. In these experiments, we marked the runtime and the number of patterns of the exact algorithm by ``---''.

Observe that the cost of the exact algorithm increases with $s_a$ and
the algorithm cannot produce any patterns within 3 minutes when $s_a$ is higher
than 0.02 on the Taxi dataset.
This is due to
the facts that (i) Taxi is a large dataset and the number of
atomic patterns compared to other datasets becomes very large as $s_a$
increases, and (ii) the Taxi network is
quite dense (e.g., denser than the MTR network), so the number of
candidate patterns at each level becomes high.
The approximate algorithm shows stable performance across all tested
$s_a$ values, with runtimes between 3-6 seconds.
This is expected, since the algorithm evaluates a fixed number $M=1$mil of candidate ODT patterns.

The weighted-approximate algorithm
verifies fewer ODT candidates compared to the unweighted version (10k
instead 1mil), so it
completes faster. On Taxi with $s_a=0.015$, the
exact algorithm requires 4.57 seconds, while the weighted method
completes in 1.00 second. On MTR, the cost
increases from 0.13 seconds at $s_a=0.015$ to 0.63 seconds at
$s_a=0.05$ for the exact algorithm.
The weighted method maintains sub-second runtime across all
configurations.
In all cases, the weighted approximate algorithm runs much faster than
the unweighted algorithm, while producing more patterns.
On MTR and Flights a significant proportion of the exact result is
detected by the approximate methods, however, their cost difference to
the exact method is not high (the approximate methods may even be
slower).
On Taxi, the weighted approximate algorithm
produces a representative sample of the patterns within a short time.
The reason behind the fact that the weighted method
is faster than the sampling approximate algorithm is that the weighted
method focuses evaluation on high-probability
candidates by ranking all possible ODT triples by their weight (based
on the number of atomic patterns each node/timeslot
participates in) and checking only the top $M$=10k candidates,
In contrast, the approximate method
randomly samples $M$=1mil candidates with equal probability.
Since patterns are not uniformly distributed in the search space, the
weighted approach evaluates fewer low-probability candidates
and finds valid patterns more efficiently.


\begin{table}[t]
\centering
\caption{Performance and pattern count for varying $s_a$ with default
  $s_r=0.5$, \linebreak $S_O=S_D=S_T=4$}
\label{tab:varying_sa}
\begin{tabular}{@{}lcccccc@{}}
\toprule
\textbf{Method} & \multicolumn{2}{c}{\textbf{Taxi Network}} & \multicolumn{2}{c}{\textbf{MTR Network}} & \multicolumn{2}{c}{\textbf{Flight Network}} \\
\cmidrule(lr){2-3} \cmidrule(lr){4-5} \cmidrule(lr){6-7}
& Time (sec) & \#patterns & Time (sec) & \#patterns & Time (sec) & \#patterns \\
\midrule
\multicolumn{7}{l}{\textit{$s_a=0.015$}} \\
Exact                 & 4.57 & 62.1k & 0.13 & 83 & 0.16 & 4 \\
Approx.\ ($M$=1mil)     & 3.08  & 24         & 2.97 & 24      & 2.92  & 1 \\
Weighted\ ($M$=10k)   & 1.00  & 55         & 0.32 & 25      & 0.81  & 2 \\
\midrule
\multicolumn{7}{l}{\textit{$s_a=0.02$}} \\
Exact                 & 7.7& 124k        & 0.18 & 210  & 0.25 & 4 \\
Approx.\ ($M$=1mil)     & 6.12  & 212       &2.97 & 52   & 3.01 & 0 \\
Weighted\ ($M$=10k)   & 1.00  & 627       & 0.33 & 53   & 0.80 & 2 \\
\midrule
\multicolumn{7}{l}{\textit{$s_a=0.05$}} \\
Exact                 & 38.1 & 715k      & 0.63 & 2{,}056 &1.20 & 60 \\
Approx.\ ($M$=1mil)     & 3.14  & 1k   & 3.08 & 788 & 3.05  & 15 \\
Weighted\ ($M$=10k)   & 0.6 & 2.19k   & 0.34 & 536    & 0.84  & 23 \\
\bottomrule
\end{tabular}
\end{table}

\subsubsection{Effect of Varying $s_r$}

Table \ref{tab:varying_sr} compares the methods for various values of
$s_r$, having
fixed $s_a$ to $0.015$ and $S_O=S_D=S_T=4$. Reducing $s_r$ increases
the cost for the exact algorithm,
as more ODT triples qualify as patterns when the support ratio
requirement is relaxed. For the Taxi Network at $s_r=0.3$,
the exact algorithm requires 11.46 seconds to find all 271k
patterns. Both approximate methods
show stable performance across all $s_r$ values, since they evaluate a fixed number of candidates.

The weighted method completes within under 1 second across all datasets
and $s_r$ values. On the MTR Network,
the exact algorithm completes in under 1 second at all tested $s_r$
values, with pattern counts ranging from 83 to 698.
The exact method runs faster than the sampling approximate method
because
the sparsity of the network greatly limits the number of generated
candidates at each level, hence, the total number of generated ODT candidates
for $S_O=S_D=S_T=4$ is relatively small.
Overall, the weighted approximate algorithm is worth using in dense
networks, where it can detect a good proportion of patterns within a
certain time bound, as opposed to the exact method, which operates
level by level and takes longer time to generate and verify (more) ODT candidates.


\begin{table}[t]
\centering
\caption{Performance and pattern count for varying $s_r$ with default $s_a=0.015$, \linebreak $S_O=S_D=S_T=4$}
\label{tab:varying_sr}
\begin{tabular}{@{}lcccccc@{}}
\toprule
\textbf{Method} & \multicolumn{2}{c}{\textbf{Taxi Network}} & \multicolumn{2}{c}{\textbf{MTR Network}} & \multicolumn{2}{c}{\textbf{Flight Network}} \\
\cmidrule(lr){2-3} \cmidrule(lr){4-5} \cmidrule(lr){6-7}
& Time (sec) & \#patterns & Time (sec) & \#patterns & Time (sec) & \#patterns \\
\midrule
\multicolumn{7}{l}{\textit{$s_r=0.3$}} \\
Exact                 &11.46 & 271k & 0.21 & 698 & 0.86 & 154 \\
Approx.\ ($M$=1mil)     & 2.93 & 227   &2.99 & 240& 2.97  & 15 \\
Weighted\ ($M$=10k)   & 0.92  & 509   & 0.32 & 206    & 0.77  & 23 \\
\midrule
\multicolumn{7}{l}{\textit{$s_r=0.4$}} \\
Exact                 &5.24   & 145k      & 0.14 & 329     & 0.30  & 4 \\
Approx.\ ($M$=1mil)     & 0.45  & 77        & 3.14 & 112     & 2.83  & 1 \\
Weighted\ ($M$=10k)   & 0.92  & 280        & 0.32 & 103     & 0.84  & 2 \\
\midrule
\multicolumn{7}{l}{\textit{$s_r=0.5$}} \\
Exact                 & 4.57 & 62.1k & 0.13 & 83 & 0.16 & 4 \\
Approx.\ ($M$=1mil)     & 3.08  & 24         & 2.97 & 24      & 2.92  & 1 \\
Weighted\ ($M$=10k)   & 1.00  & 55         & 0.32 & 25      & 0.81  & 2 \\
\bottomrule
\end{tabular}
\end{table}

\subsubsection{Effect of Varying $S_O, S_D, S_T$}

Table \ref{tab:varying_odt} reports runtime and pattern count for
different values of the sizes $(S_O,S_D,S_T)$
of the pattern components. In all experiments, we fixed $s_a=0.015$
and $s_r=0.5$.

As we increase
the bounds from (3,3,3) to (4,4,4), we observe how the computational cost and pattern counts change.
For the smaller configurations with $S_T=3$, the exact algorithm shows
stable performance on the Taxi Network, completing in 12.21
seconds at (3,3,3) and finding 271,444 patterns. As we increase the
time dimension to $S_T=4$ and $S_T=5$ while keeping $S_O=S_D=3$, the
exact algorithm's runtime increases from 12.21 seconds to 24.46
seconds on Taxi, while pattern counts remain at similar levels
(271k to 300k). The approximate method scales from 3.16 seconds
to 4.30 seconds across these configurations,
discovering between 210 and 244 patterns. The weighted method
maintains consistent sub-second performance,
completing in around 1 second across all configurations from (3,3,3)
to (3,3,5), while finding 2 to 3 times more patterns
than the unweighted approximate method.

At the (4,4,4) configuration, the exact algorithm requires 4.57
seconds on the Taxi Network and finds 62.1k patterns.
This shows a significant reduction in pattern count compared to the
(3,3,3) configuration, which is expected as larger
patterns impose stricter structural constraints. The approximate
method requires 3.08 seconds and finds only 24 patterns,
while the weighted method completes in 1.00 second and finds 55
patterns. On the MTR Network at (4,4,4), the exact
algorithm completes in only 0.13 seconds finding 83 patterns, demonstrating the efficiency gains on sparser networks.

At the largest configuration (5,5,4), the exact algorithm times out on
all three datasets.
The approximate method requires 8-9 seconds, while the weighted method
maintains around 1 second runtime.
On the Taxi Network at (5,5,4), the approximate method finds 8,677
patterns while the weighted method finds 8,981 patterns,
showing that at very large pattern sizes both methods discover
substantial numbers of patterns. On the MTR Network at (5,5,4),
the approximate method finds 6,812 patterns while the weighted method
finds 1,870 patterns. On the Flight Network, pattern
counts are low across all configurations.

\begin{table}[t]
\centering
\caption{Performance and pattern count for different combinations of O, D, T bounds with $s_a=0.015$, $s_r=0.5$}
\label{tab:varying_odt}
\begin{tabular}{@{}lcccccc@{}}
\toprule
\textbf{Method} & \multicolumn{2}{c}{\textbf{Taxi Network}} & \multicolumn{2}{c}{\textbf{MTR Network}} & \multicolumn{2}{c}{\textbf{Flight Network}} \\
\cmidrule(lr){2-3} \cmidrule(lr){4-5} \cmidrule(lr){6-7}
& Time (sec) & \#patterns & Time (sec) & \#patterns & Time (sec) & \#patterns \\
\midrule
\multicolumn{7}{l}{\textit{$S_O$=3, $S_D$=3, $S_T$=3}} \\
Exact                 & 12.21 & 271k & 0.21 & 698   & 0.85 & 154 \\
Approx.\ ($M$=1mil)     & 3.16  & 210       & 3.15 & 249   & 2.95 & 11  \\
Weighted\ ($N$=10K)   & 1.02  & 509       & 0.33 & 206   & 0.81 & 23  \\
\midrule
\multicolumn{7}{l}{\textit{$S_O$=3, $S_D$=3, $S_T$=4}} \\
Exact                 & 17.69 & 300k & 0.28 & 733   & 1.08 & 44  \\
Approx.\ ($M$=1mil)     & 3.77  & 244       & 3.58 & 271   & 3.37 & 1   \\
Weighted\ ($N$=10K)   & 0.94  & 557       & 0.32 & 230   & 0.83 & 4   \\
\midrule
\multicolumn{7}{l}{\textit{$S_O$=3, $S_D$=3, $S_T$=5}} \\
Exact                 & 24.46 & 280k & 0.35 & 606   & 1.27 & 8   \\
Approx.\ ($M$=1mil)     & 4.30  & 214       & 4.10 & 222   & 4.06 & 2   \\
Weighted\ ($N$=10K)   & 0.94  & 518       & 0.33 & 201   & 0.81 & 4   \\
\midrule
\multicolumn{7}{l}{\textit{$S_O$=4, $S_D$=4, $S_T$=4}} \\
Exact                 & 4.57 & 62.1k & 0.13 & 83 & 0.16 & 4 \\
Approx.\ ($M$=1mil)     & 3.08  & 24         & 2.97 & 24      & 2.92  & 1 \\
Weighted\ ($N$=10K)   & 1.00  & 55         & 0.32 & 25      & 0.81  & 2 \\
\midrule
\multicolumn{7}{l}{\textit{$S_O$=5, $S_D$=5, $S_T$=4}} \\
Exact                 &---  & ---      & ---  & ---   & ---  & --- \\
Approx.\ ($M$=1mil)     & 8.44    & 8{,}677  & 7.72    & 6.8k & 8.96   & 0   \\
Weighted\ ($N$=10K)   & 1.14    & 8{,}981  & 0.38    & 1.8k & 0.81   & 0   \\
\bottomrule
\end{tabular}
\end{table}

Table \ref{tab:varying_odt_compact} summarizes the runtime
comparison across
three representative configurations. The results show that
the approximate methods are not worth applying when the pattern
size parameters  $(S_O,S_D,S_T)$ are small or when the network is
sparse (MTR).
On the other hand on dense networks (Taxi) or for large  pattern
size parameters, the weighted approximate algorithm is consistently
superior compared to the exact method in terms of runtime and it is
worth applying. 
When comparing the two versions of the approximate algorithm, the
weighted method prevails as it completes faster (typically 3-8 times), while finding 
a similar number of patterns compared to the unweighted approximate method.

\begin{table}[t]
\centering
\caption{Runtime comparison (in seconds) for different O, D, T configurations with $s_a=0.015$, $s_r=0.5$}
\label{tab:varying_odt_compact}
\begin{tabular}{@{}l@{~}c@{~}c@{~}c@{~}c@{~}c@{~}c@{~}c@{~}c@{~}c@{}}
\toprule
& \multicolumn{3}{c}{\textbf{$S_O$=3, $S_D$=3, $S_T$=3}} & \multicolumn{3}{c}{\textbf{$S_O$=4, $S_D$=4, $S_T$=4}} & \multicolumn{3}{c}{\textbf{$S_O$=5, $S_D$=5, $S_T$=4}} \\
\cmidrule(lr){2-4} \cmidrule(lr){5-7} \cmidrule(lr){8-10}
\textbf{Dataset} & Exact & Approx. & Weighted & Exact & Approx. & Weighted & Exact & Approx. & Weighted \\
\midrule
Taxi & 12.21 & 3.16 & 1.02 & 4.57 & 3.08 & 1.00 & timeout & 8.44 & 1.14 \\
MTR & 0.21 & 3.15 & 0.33 & 0.13 & 2.97 & 0.32 & timeout & 7.72 & 0.38 \\
Flight & 0.85 & 2.95 & 0.81 & 0.16 & 2.92 & 0.81 & timeout & 8.96 & 0.81 \\
\bottomrule
\end{tabular}
\end{table}

Overall, the exact algorithm is limited by its high computational cost on
dense networks like Taxi,
particularly at large pattern sizes. The approximate methods provide
stable, predictable performance.
The choice of algorithm depends on completeness requirements: the
exact algorithm guarantees finding all patterns when
it can complete, while the approximate methods enable discovery of
large patterns in cases where the exact method times out.

\subsection{Use cases}
Lastly, we explore the use of ODT patterns in real-world
applications. To do that, we restricted the origin and time
dimensions, according to Section \ref{sec:restricted}, and identified
the most popular destinations for different times of the day. In the
first experiment, we used the Taxi dataset.
Table \ref{table:usecase2} shows some of these patterns. We first
restricted O to be Financial District, Manhattan and T to morning,
lunch, and afternoon timeslots. This gave us the most popular
destinations during these timeslots, extented region 
Greenwich Village South and North, and Tribeca, extended
region Greenwich Village South and North, Tribeca and Battery Park, and
extented region MidTownEast and Centre, Murray Hill, Union Square, and
Stuy Town. Our study shows that people tend to move toward residential
areas or popular dining spots like Union Square during the afternoon,
suggesting a preference for destinations near their homes by the end
of the day. An interesting observation is the shift in movement
across districts during lunch, possibly due to limited dining
options in the Financial District. Insights like these could help
restaurant owners identify ideal locations for new establishments,
considering the demand in these areas. 

\begin{table}[ht]
\caption{Use case - Taxi Dataset}
\centering
\tiny
\begin{tabular}{|c|c|c|}
 \hline
 Origin &Timeslots & popular destinations\\
 \hline
   \hline
 FinancialDistrict&[8:30-9:30]&GreenwichVillageS, GreenwichVillageN \\
 FinancialDistrict&[8:30-9:30]&GreenwichVillageS, GreenwichVillageN, Tribeca\\
 \hline
  FinancialDistrict&[12:30-14:00]&GreenwichVillageS, GreenwichVillageN,
                                  Tribeca\\
  FinancialDistrict&[12:30-14:00]&GreenwichVillageS, GreenwichVillageN,
                                  Tribeca, BatteryPark\\
  \hline
 FinancialDistrict&[17:30-19:30]&MidTownCentre, MidTownEast, MurrayHill\\  
FinancialDistrict&[17:30-19:30]&MidTownCentre, MidTownEast,
                                 MurrayHill, Union
  Square, StuyTown\\[0.2ex]
  \hline
\end{tabular}
 \label{table:usecase2}
 \end{table}

We repeated the experiment on the MTR data to obtain some interesting
patterns, such as those shown in Table \ref{table:usecase1} considering
different origin points.
As representative patterns, we show popular destinations for people
who move from popular regions such as Mong Kok in morning peak hours.
The extracted patterns show that the most popular destinations are
an extended region covering the center of Hong Kong (e.g., Central) and extended regions which 
includes the city center and destinations from the city center along the
north coast of Hong Kong island.
In case of an service disruption incident
at the origin,
the transportation company can
use our approach to extract constrained patterns to
predict the travel needs of affected passengers and schedule emergency
transportation for them (e.g., buses).

 \begin{table}[ht]
 \caption{Use case - MTR Dataset}
\centering
  \scriptsize
 \begin{tabular}{|c|c|c|}
 \hline
 Origin &Timeslots & popular destinations\\
 \hline
 \hline
 Mong Kok&[8:30-9:30]& [Central-Admiralty]\\
 Mong Kok &[8:30-9:30]&[Central-Admiralty],[Exhibition Centre-University] \\
 \hline
 Wan Chai &[17:30-18:30]&[North Point-Quarry Bay]\\
 Wan Chai &[17:30-18:30]&[North Point-Quarry Bay][Yau Tong-Kwun Tong]\\[0.2ex]  
 \hline
 \end{tabular}
 \label{table:usecase1}
 \end{table}

\section{Conclusions}\label{sec:conc}
In this paper, we have studied the problem of enumeration
origin-destination-timeslot (ODT) patterns at varying granularity from
a database of trips. To our knowledge, this is the first work that
formulates and studies this problem. Due to the huge number of
region-time combinations that can formulate a candidate pattern, the
problem is hard. We explore the problem space level-by-level, buiding
on a weak monotonicity property of patterns. We propose a number of
optimizations that greatly reduce the cost of the baseline pattern
enumeration algorithm. To eliminate the possibly huge number of ODT
patterns, which take too long to enumerate and analyze, we propose
practical variants of the mining problem, where we restrict the size
of patterns and/or the regions/timeslots included in them. In
addition, we suggest an interesting definition of rank-based patterns
and we study their efficient enumeration.
Finally, we propose an approximate algorithm and a weighted version of
it, which can detect representative patterns of a given size much
faster than the exact algorithm.  
Experiments with three real
datasets demostrate the effectiveness of the proposed techniques. In
the future, we plan to study database queries that apply on the trips
table, to analyze the outflow (or inflow) of generalized regions to
(or from) other regions at specific time intervals. Such a tool,
together with our pattern enumeration tools can assist transportation
analysts or companies in planning or restructuring.  

\backmatter

\bibliography{references}

\end{document}